\documentclass[preprint2]{aastex}

\shorttitle{Observations of Comet 103P/Hartley 2}
\shortauthors{Gicquel et al.}

\begin{document}


\title{Ground-based Multiwavelength Observations of Comet 103P/Hartley 2}

\author{A. Gicquel \altaffilmark{1,2}, S. N. Milam \altaffilmark{2}, G. L. Villanueva \altaffilmark{1,2}, A. J. Remijan \altaffilmark{3}, I. M. Coulson \altaffilmark{4}, Y.-L. Chuang \altaffilmark{5}, S. B. Charnley \altaffilmark{2}, M. A. Cordiner \altaffilmark{1,2} and Y.-J. Kuan \altaffilmark{5,6}}

\altaffiltext{1}{Catholic University of America, Physics Department, 620 Michigan ave NE, Washington, DC;}
\altaffiltext{2}{Goddard Center for Astrobiology, NASA Goddard Space Flight Center, 8800 Greenbelt Rd., Greenbelt, MD 20771, USA; adeline.gicquel@nasa.gov, stefanie.n.milam@masa.gov, geronimo.l.villanueva@nasa.gov, steven.b.charnley@nasa.gov, martin.a.cordiner@nasa.gov.}
\altaffiltext{3}{National Radio Astronomy Observatory, 520 Edgemont Road, Charlottesville, VA 22903, USA; aremijan@nrao.edu.}
\altaffiltext{4}{Joint Astronomy Centre, 660 N. A'ohoku Place University Park, Hilo, Hawaii 96720, USA; i.coulson@jach.hawaii.edu}
\altaffiltext{5}{National Taiwan Normal University, 88 Sec. 4 Ting-Chou Rd., Taipei 116, Taiwan; ylchuang@std.ntnu.edu.tz}
\altaffiltext{6}{Academia Sinica Institute of Astronomy $\&$ Astrophysics (ASIAA), Taipei 106, Taiwan; kuan@ntnu.edu.tw}

\begin{abstract}
The Jupiter-family comet 103P/Hartley 2 (103P) was the target of the NASA EPOXI mission. In support of this mission, we conducted observations from radio to submillimeter wavelengths of comet 103P in the three weeks preceding the spacecraft rendezvous on UT 2010 November 4.58. This time period included the passage at perihelion and the closest approach of the comet to the Earth. Here we report detections of HCN, $\rm H_2CO$, CS, and OH and upper limits for HNC and DCN towards 103P, using the Arizona Radio Observatory Kitt Peak 12m telescope (ARO 12m) and submillimeter telescope (SMT), the James Clerk Maxwell Telescope (JCMT) and the Greenbank Telescope (GBT). The water production rate, $ Q_{H_2O}$ = (0.67 - 1.07) $\times$ $10^{28}$ $\rm s^{-1}$, was determined from the GBT OH data. From the average abundance ratios of HCN and $\rm H_2CO$ relative to water (0.13 $\pm$ 0.03 $\%$ and 0.14 $\pm$ 0.03 $\%$, respectively), we conclude that $\rm H_2CO$ is depleted and HCN is normal with respect to typically-observed cometary mixing ratios. However, the abundance ratio of HCN with water shows a large diversity with time. Using the JCMT data we measured an upper limit for the DCN/HCN ratio $<$0.01. Consecutive observations of $ortho$-$\rm H_2CO$ and $para$-$\rm H_2CO$ on November 2 (from data obtained at the JCMT), allowed us to derive an $ortho:para$ ratio (OPR) $\approx$ 2.12 $\pm$ 0.59 (1$\sigma$), corresponding to $T_{spin}$ $>$ 8 K (2$\sigma$).
\end{abstract}


\section{Introduction}
Comets are probably the least-altered bodies in the Solar System \citep{Festou_2004}. As such, they can provide key insights into physical and chemical processes occurring during its origin and earliest evolutionary epochs, including the origin of long-period and short period comets \citep{Duncan_2004} and the formation and composition of planets \citep{DR_2009, Bast_2013}. By studying comets from different reservoirs we can probe the different environments in which they formed and also better understand their role in initiating prebiotic chemistry on the early Earth through delivery of water and organic matter by cometary impacts \citep{Chyba_1990}. 
Comets are primarily located in two distinct reservoirs of the Solar System: the Oort Cloud and the Kuiper Belt. The Oort Cloud is a source of long-period comets and (probably) Halley-type comets \citep{Levison_2001}. The Kuiper Belt is a source of Jupiter family comets \citep{Duncan_2004, Levison_1997}. To explore the connections among the formation environment, subsequent chemical evolution, and present composition of comets, efforts have been made to develop chemical taxonomies \citep{AHearn_1995, DBM_2004, DBM_2005, Feldman_2005, Crovisier_2009, Fink_2009, Mumma_C_2011}.\\
Important cosmogonic quantities for the investigation of primordial constitutions (such as the $ortho:para$ and D/H ratios), can be obtained by observations of comets from ground-based, space-based and airborne observatories \citep{Crovisier_1997, Biver_2007, Hartogh_2011, Reach_2013, Hines_2014}, as well as from laboratory analyses of cometary (interplanetary) dust particles collected in the stratosphere \citep{Starkey_2013}. These data provide measurements of chemical composition, isotopic ratios and molecular spin ratios, which are important to help establish the contribution of the Solar System's natal molecular cloud to the compositions of primitive materials and comets. In addition, isotopic ratios such as D/H can be compared with the value for the Earth's ocean water to investigate the possible importance of cometary water delivery.

Comet rendezvous space missions to individual comets present unique scientific opportunities. $In$ $situ$ spacecraft observations of several comets have allowed their nuclei to be imaged and the gas and dust properties of their comae to be studied in great detail. These include: 1P/Halley \citep{Newburn_1991}; 81P/Wild 2 \citep{Brownlee_2006}; 9P/Tempel \citep{AHearn_2005}; 103P/Hartley 2 \citep{Ahearn_2011}; and the current ROSETTA mission to 67P/Churyumov-Gerasimenko \citep{Schulz_2012}. Ground-based observing campaigns have also played an important role in support of space missions \citep{Meech_2005, Meech_2011, Knight_2007} and, in the case of 81P/Wild 2, laboratory analyses of the returned Stardust samples have yielded unique insights into the nature of cometary dust \citep{McKeegan_2006, Clemett_2010}.

Comet 103P/Hartley 2 (hereafter 103P) is a Jupiter-family comet, with a short orbital period (6.5 years), and a low orbital inclination. It was the target of the NASA EPOXI mission \citep{Ahearn_2011, Meech_2011}. The comet passed perihelion on 2010 October 28 at $R_h$ = 1.059 AU and on 2010 October 21 made an exceptionally close approach to Earth at $\Delta$ = 0.12 AU. In support of the EPOXI mission, we conducted observations at radio and submillimeter wavelengths of comet 103P in the three weeks preceding the spacecraft rendezvous on UT 2010 November 4.58.
Here we report detections of HCN, $\rm H_2CO$, CS, and OH and upper limits for HNC and DCN, using the 12m Arizona Radio Observatory Kitt Peak (12m) and the 10m submillimeter telescope (SMT), as well as the 15m James Clerk Maxwell Telescope (JCMT) and the 100m Robert C. Byrd Greenbank Telescope (GBT). We present the observational results and determine physical parameters such as column densities, and production rates. Finally, the $ortho:para$ ratio has been derived from $\rm H_2CO$, and an upper limit on the D/H ratio was obtained from DCN and HCN measurements.

\section{Observations}
Comet 103P/Hartley 2 was discovered on 1984 June 4 by Malcolm Hartley at the Siding Spring Observatory \citep{Hartley_1986}. 103P has been frequently observed over the 30 years following its discovery, both by ground-based and space telescopes. Observations provided information about the gas production rate \citep{AHearn_1995, Crovisier_1999, Colangeli_1999, Fink_2009, Combi_2011}, and the mean radius \citep{Lisse_2009}. All the observations presented here adopted the same JPL/Horizons (Jet Propulsion Laboratory) ephemeris number 183.

\subsection{ Green Bank Telescope (GBT)}
Observations of comet 103P were made between 2010 October 13 and October 31 for monitoring emission from the OH: J = 3/2, $\Omega$=3/2, F=$\rm 1^+$-$\rm1^-$ (1667.3 MHz/18 cm) lines at L-band (Program ID GBT/10C-059). A barycentric velocity reference frame was assumed for all observations. The line of sight radial velocity on each day recorded by the JPL ephemeris was then subtracted from the data to set each spectral feature to a 0 km $\rm s^{-1}$ cometocentric velocity. The GBT spectrometer was configured in its eight intermediate-frequency (IF), 12.5 MHz three-level mode, which enabled observing four 12.5 MHz frequency bands simultaneously in two polarizations through the use of offset oscillators in the IF. Antenna temperatures were recorded on the $T_{A}^{*}$ scale \citep{Ulich_1976} with estimated 20$\%$ uncertainties. Data were taken in the OFF-ON position-switching mode, with the OFF position 60' east in azimuth with respect to the ON-source position. A single scan consisted of 2 minutes in the OFF-source position followed by 2 minutes in the ON-source position. Automatically-updated dynamic pointing corrections were employed based on real-time temperature measurements of the structure input to a thermal model of the GBT. Zero points were adjusted at the beginning of each observing run by pointing on a nearby quasar. The pointing accuracy of the GBT is 5'' (rms) for blind pointing and 7'' (rms) for offset pointing. The GBT beam size at this frequency is $\approx$ 8'.

\subsection{Arizona Radio Observatory 12m (12m) and Submillimeter Telescope (SMT)}
Observations of HCN: J = 3-2 (265.8864 GHz), $o$-$\rm H_2CO$: $\rm J_{K_a,K_c}$ = $\rm 3_{1,2}$-$\rm 2_{1,1}$ (225.6978GHz), HCN: J = 1-0 (88.6318GHz), HCN: J = 2-1 (177.2612GHz), and CS: J = 4-3 (97.9809GHz), toward comet 103P were taken between 2010 October 22 and November 4 using the facilities of the Arizona Radio Observatory (ARO): the 12m telescope on Kitt Peak, Arizona and the SMT on Mount Graham, Arizona. The 1 mm observations were carried out at the SMT with a dual-polarization ALMA Band 6 receiver system, employing sideband-separating mixers with an image rejection of typically 15--20 dB. The backends employed were a 2048 channel 1 MHz filter bank used in parallel (2 $\times$ 1024) mode, and 250 kHz filter also in parallel (2 $\times$ 250). The temperature scale at the SMT is $T_{A}^{*}$; radiation temperature is then defined as $T_R$ = $T_{A}^{*}$/$\eta_b$, where $\eta_b$ is the main beam efficiency. The 2 and 3mm observations were conducted at the ARO 12 m using dual-polarization SIS mixers, operated in single-sideband mode with the image rejection $\geq $ 20 dB. Filter banks with 512 channels of 100 and 250 KHz resolutions were used simultaneously in parallel mode for the measurements, along with an autocorrelator with 782 kHz resolution. The intensity scale of the 12 m is the chopper-wheel corrected antenna temperature, $T_{R}^{*}$, including forward spillover losses, which is converted to radiation temperature by $T_R$ = $T_{R}^{*}$/$\eta_c$, where $\eta_c$ is the corrected beam efficiency. Data at both facilities were taken in position-switching mode with an off position 30' west in azimuth. The JPL ephemeris was used to determine the cometary position using the orbital elements. The pointing accuracy is estimated at 1'' rms for the SMT and 5'' for the ARO 12m. Focus and positional accuracy were checked periodically on nearby planets or masers.

\subsection{James Clerk Maxwell Telescope (JCMT)}
Sub-millimeter spectroscopic observations of 103P (Programme ID m10bu13) were made from the JCMT, located at the 4000m level on Mauna Kea, Hawaii, on UT 2010 October 21, 22, 23, 25 and November 02, 03, 04, using the HARP heterodyne array. HARP is a 4 $\times$ 4 array receiver, but was used here in a single-receptor mode. HARP is remotely and quickly tunable to frequencies in the 325-375GHz ranges and utilizes an image sideband suppressor. The output from the HARP receiver is fed to the `ACSIS' digital autocorrelation spectrometer. For this work, ACSIS was configured at its highest frequency resolutions (31MHz). Observations of HCN: J = 4-3 (354.5055GHz), HCN: J = 3-2 (265.8864GHz), $p$-$\rm H_2CO$: $\rm J_{K_a,K_c}$ = $\rm 5_{0,5}$-$\rm 4_{0,4}$ (362.7360GHz), $o$-$\rm H_2CO$: $\rm J_{K_a,K_c}$=$\rm 5_{1,5}$-$\rm 4_{1,4}$ (351.7686GHz), DCN: J = 5-4 (362.0465GHz) and HNC: J = 4-3 (362.6303GHz) were measured at the JCMT.

At these frequencies, the JCMT has a Gaussian beam of size $\approx$ 15'' (full width at half power). The HARP receptors are separated by about 30'', and so the emission from the comet is anticipated to be encompassed by a single receptor -- the `pointing' receptor. Pointing of the telescope is achieved by doing spectral-line `five-points' on nearby line sources, and pointing accuracy is estimated at 2" rms in each of two orthogonal coordinates (azimuth and elevation). A pointing check was performed approximately every hour. Telescope focus is similarly maintained throughout the night by measures of bright line-sources. Calibration of the ($T_A^*$) brightness scale is achieved by making measures at standard frequencies (eg CO: J = 3-2, 345GHz) of astronomical sources used as reference calibrators. 

The opacity of the sky above JCMT was measured by a water vapor meter (WVM) mounted so as to measure along the telescope line-of-sight. Opacity is expressed as if measured at the zenith at 225GHz. The opacity (in nepers) on UT Oct 21, 22, 23, 25 was, on average, 0.10, 0.05, 0.05, 0.05 and on Nov 02, 03, 04 was 0.09, 0.11, 0.15. The latter conditions made observations difficult at these HARP (B-band) frequencies, but conditions otherwise were most suitable for this program.

\section{Results}

Observed lines parameters including UT dates, observing frequencies ($\nu$), beam size ($\Theta_b$), diameter of the projected beam size on the comet (D), beam efficiency ($\eta_c$ or $\eta_b$), temperature ($T_R^{*}$ or $T_A^{*}$), FWHM line width ($\Delta v_{1/2} $), integrated intensity ($\int T_R^{*}$ $\Delta v_{1/2}$), heliocentric distance ($R_h$) and comet distance at the times of measurements ($\Delta$) are listed in Table \ref{Obs_parameters}. The uncertainties of the line width and the integrated line intensity are determined from a Gaussian fit and are reported at 1$\sigma$ in the Table \ref{Obs_parameters}.

\begin{figure}[h!]
\centering
\includegraphics[scale=0.35]{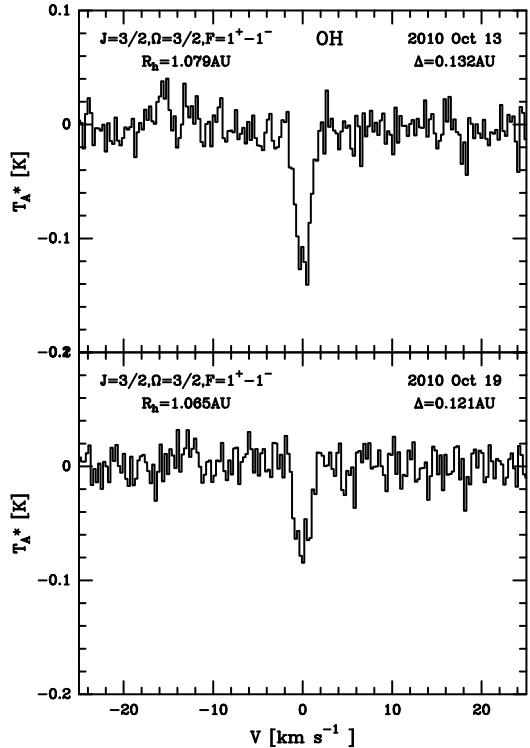}
\caption{Detection of OH taken on 2010 October 13 and 19 with the GBT. The spectral resolution is $\approx$ 0.274 km $\rm s^{-1}$. Additional spectra from the observing campaign for OH is available in Appendix \ref{Appendix_OH}. }
\label{GBT_OH}
\end{figure}

\begin{figure}[h!]
\centering
\includegraphics[scale=0.28]{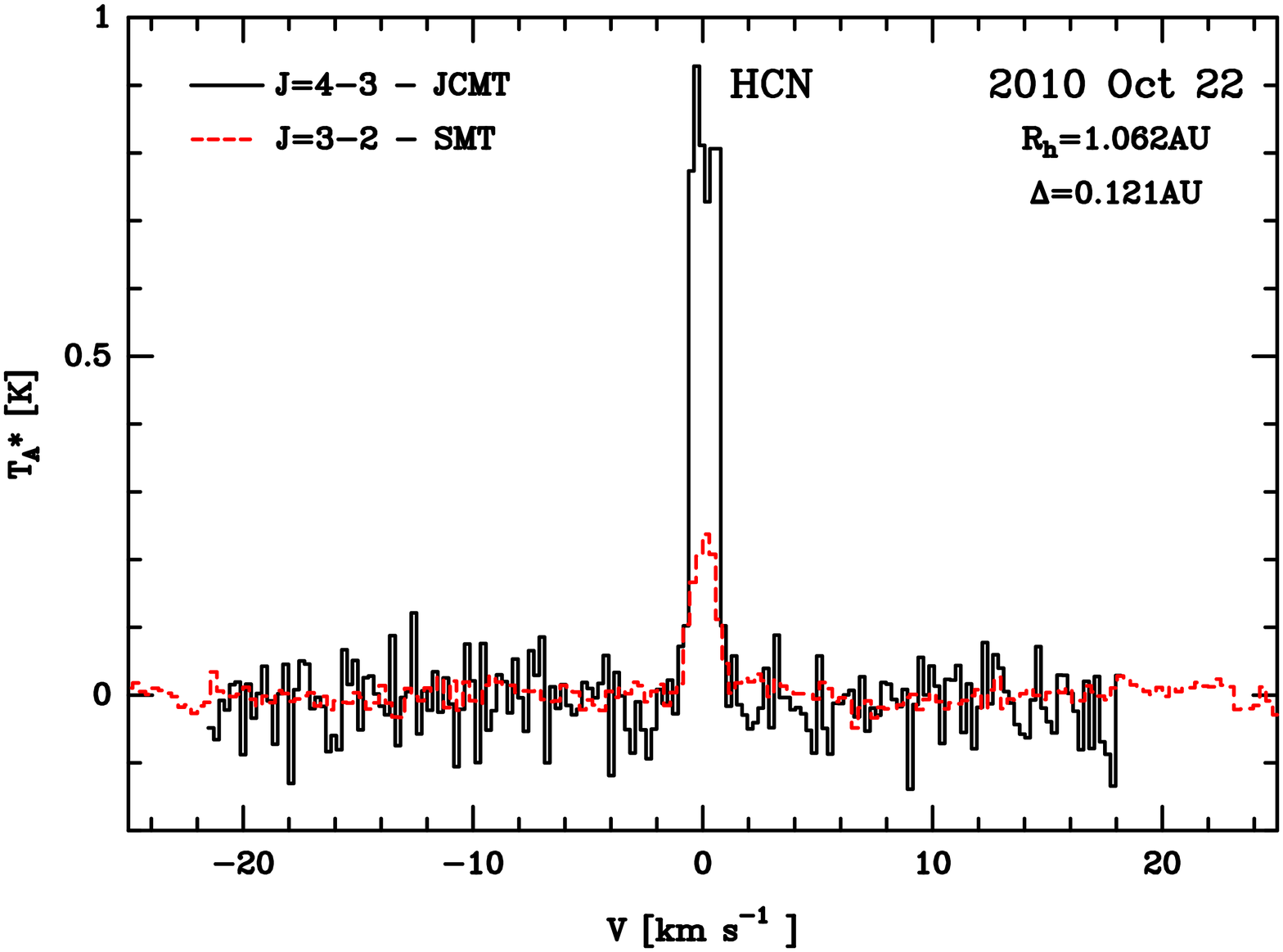}
\caption{Detection of HCN taken on 2010 October 22 with the JCMT (dark solid line) and the SMT (red dashed line). The spectral resolution are $\approx$ 0.232 km $\rm s^{-1}$ and $\approx$ 0.280 km $\rm s^{-1}$ for the JCMT and the SMT, respectively. (For interpretation of the references to color in this Figure legend, the reader is referred to the web version of this article.). The detections of HCN from all facilities and dates are available in Appendix \ref{Appendix_HCN}. }
\label{HCN_data}
\end{figure}

\begin{figure}[h!]
\centering
\includegraphics[scale=0.28]{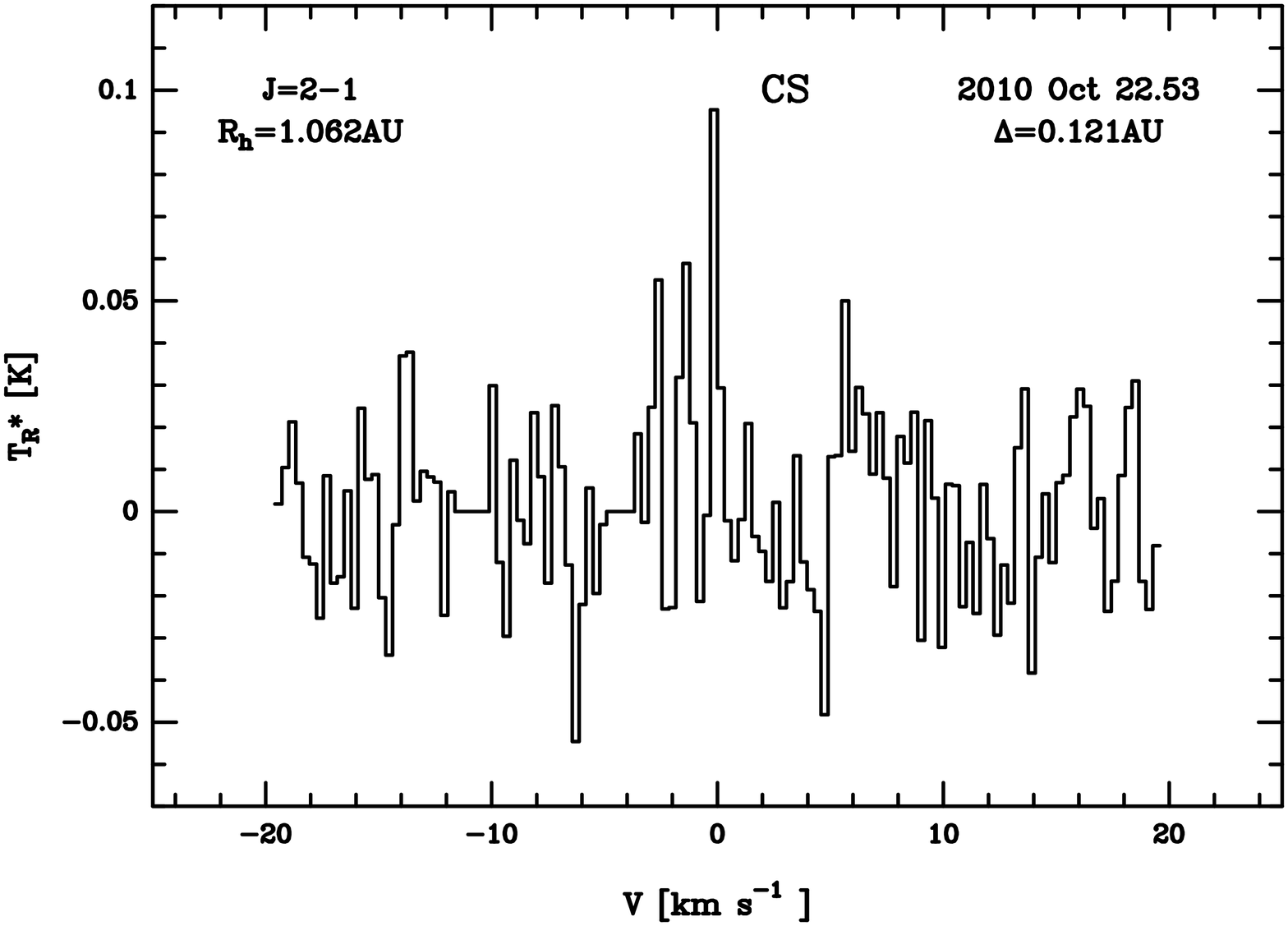}
\caption{Detection of CS 2-1 taken on 2010 October 22 with the 12m. The spectral resolution is $\approx$ 0.310 km $\rm s^{-1}$.}
\label{CS_12m}
\end{figure}

\begin{figure}[h!]
\centering
\includegraphics[scale=0.35]{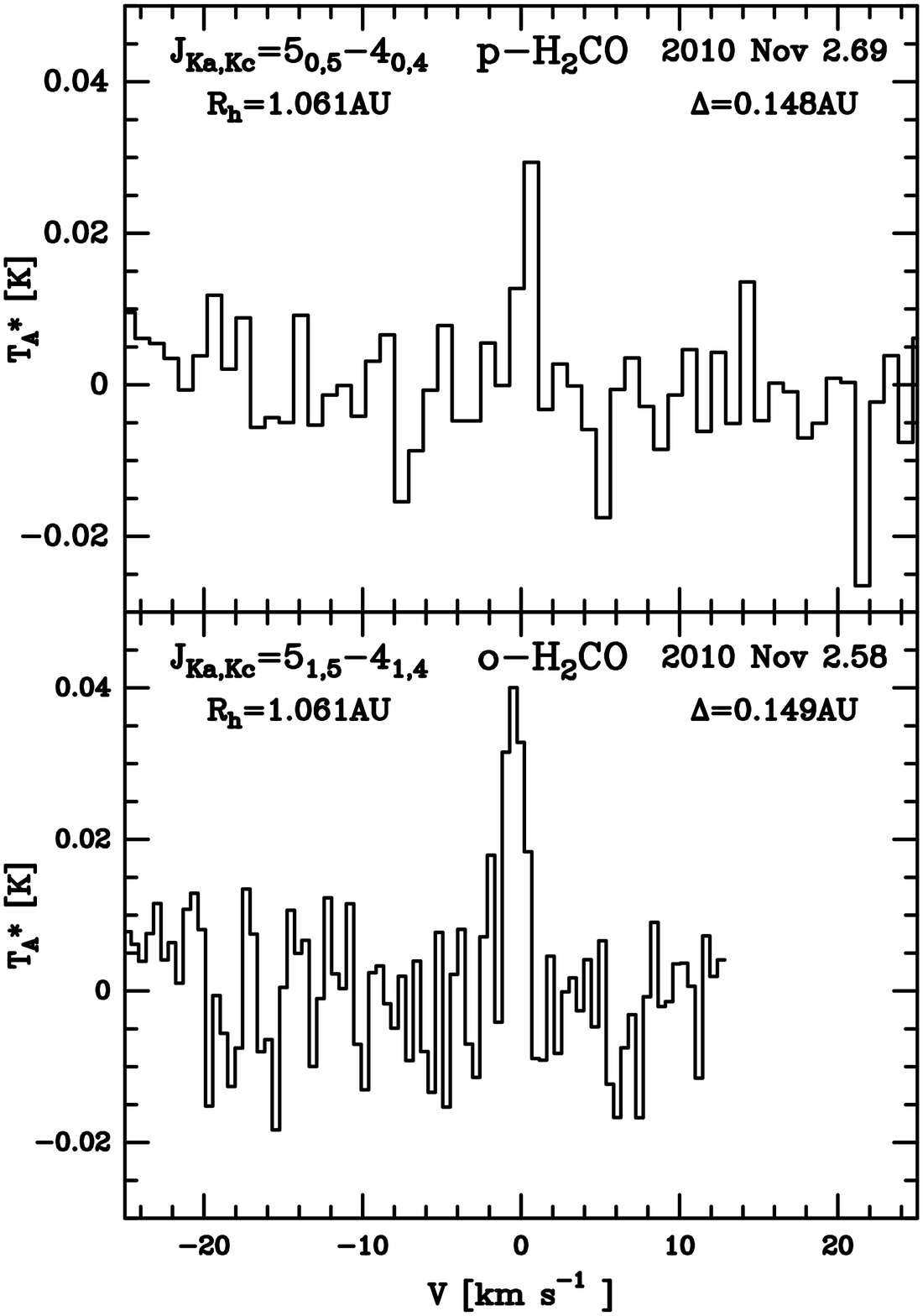}
\caption{Detection of $\rm H_2CO$ taken on 2010 November 2 with the JCMT. The spectral resolution are $\approx$ 0.909 km $\rm s^{-1}$ and $\approx$ 0.468 km $\rm s^{-1}$ for $p$-$\rm H_2CO$ and $o$-$\rm H_2CO$, respectively. The detection of $\rm H_2CO$ from all facilities and dates are available in Appendix \ref{Appendix_H2CO}.}
\label{H2CO_data}
\end{figure}

Representative spectra are shown in Figures \ref{GBT_OH} -- \ref{H2CO_data} from all facilities. All the spectra are plotted in the cometocentric velocity frame. The UT date of each observed transition is located in the Figure. All data are included as supplemental information.

Figure \ref{GBT_OH} displays the OH detection toward comet 103P/Hartley 2 from the GBT taken on 2010 October 13 and 19. The GBT data were measured in a monitoring sequence every 1, 2, 3 or 5 days. The OH line inversion and gas production rates on these days are reported in Table \ref{OH_Results}. 

Figure \ref{HCN_data} displays the J = 4-3 and the J = 3-2 transitions of HNC on 2010 October 22 toward comet 103P/Hartley 2 from the JCMT and SMT, respectively. We detected four transitions of HCN with three telescopes (SMT, 12m and JCMT). Gas production rates from these transition are reported in Table \ref{Qgas}. Figure \ref{HCN_data} shows two transitions observed on the same day. Furthermore, the line is mostly resolved with the JCMT and partially with the SMT.

Figure \ref{CS_12m} displays the J = 2-1 transitions of CS on 2010 October 22 toward comet 103P/Hartley 2 from the 12m. Gas production rates from this transition are reported in Table \ref{Qgas}. These data were collected with a resolution of $\approx$ 0.310 km $\rm s^{-1}$ and show a narrow line width. The line profile is not resolved here and only provides details on abundance and not molecular origin within the coma.

Figure \ref{H2CO_data} displays respectively the $J_{K_a,K_c}$ = $\rm 5_{0,5}$-$\rm 4_{0,4}$ and $\rm J_{K_a,K_c}$ = $\rm 5_{1,5}$-$\rm 4_{1,4}$ transitions of $p$-$\rm H_2CO$ and $o$-$\rm H_2CO$ on 2010 November 2 toward comet 103P/Hartley 2 from the JCMT. We observed $p$-$\rm H_2CO$ and $o$-$\rm H_2CO$ the same day only with the JCMT. A detailed discussion on the $ortho:para$ ratio can be found in section \ref{SECTION_OPR}. 

\begin{deluxetable}{lllllllllllll}
\tabletypesize{\tiny}
\rotate
\tablecaption{Observations of OH, HCN, $\rm H_2CO$, CS, DCN and HNC with the GBT, JCMT, SMT and 12m toward comet 103P/Hartley 2. a. $T_R^{*}$ and $\eta_c$ for the 12m, $T_A^{*}$ and $\eta_b$ for the SMT, JCMT and GBT.\label{Obs_parameters}}
\tablewidth{0pt}
\tablehead{
\colhead{Line} & \colhead{Telescope} & \colhead{ Transition } & \colhead{UT Date } & \colhead{$\nu$} &
\colhead{$\Theta_b$} & \colhead{D} & \colhead{$\eta_c$ or} &
\colhead{$T_R^{*}$ or $T_A^{*}$$^A$} & \colhead{$\Delta v_{1/2}$} &
\colhead{$\int T_R^{*}$ $\Delta v_{1/2}$} & \colhead{$\rm R_h$} & \colhead{$\Delta$} \\

\colhead{} & \colhead{} & \colhead{} & \colhead{ } & \colhead{(MHz)} &
\colhead{('')} & \colhead{(km)} & \colhead{$\eta_b$$^A$} &
\colhead{(K) } & \colhead{(km $s^{-1}$)} &
\colhead{(K km $s^{-1}$)} & \colhead{(AU)} & \colhead{(AU)} 
}
\startdata
OH &GBT&multiple lines &2010 Oct 13.04& 1667.3& 452.8& 43348& 0.95& -0.13 $\pm$ 0.01& 1.84 $\pm$ 0.13& -0.24 $\pm$ 0.02& 1.079& 0.132 \\
& & &2010 Oct 15.04& & & 41706& & -0.14 $\pm$ 0.01& 2.05 $\pm$ 0.14& -0.28 $\pm$ 0.02& 1.074& 0.127 \\
& & &2010 Oct 16.40& & & 40721& & -0.10 $\pm$ 0.01& 1.74 $\pm$ 0.15& -0.18 $\pm$ 0.03& 1.071& 0.124 \\
& & &2010 Oct 19.44& & & 39736& & -0.07 $\pm$ 0.01& 1.65 $\pm$ 0.21& -0.12 $\pm$ 0.02& 1.065& 0.121 \\
& & &2010 Oct 22.31& & & 39736& & -0.03 $\pm$ 0.01& 2.09 $\pm$ 0.28& $<$ -0.07 & 1.062& 0.121 \\
& & &2010 Oct 23.46& & & 40064& & -0.04 $\pm$ 0.01& 1.85 $\pm$ 0.39& $<$ -0.07 & 1.061& 0.122 \\
& & &2010 Oct 24.46& & & 40393& & -0.03 $\pm$ 0.01& 1.08 $\pm$ 0.32& $<$ -0.03 & 1.060& 0.123 \\
& & &2010 Oct 26.44& & & 41706& & 0.01 $\pm$ 0.01& 1.86 $\pm$ 0.75& $<$ 0.03 & 1.059& 0.127 \\
& & &2010 Oct 28.18& & & 43020& & 0.03 $\pm$ 0.01& 1.44 $\pm$ 0.34& $<$ 0.04 & 1.059& 0.131 \\
& & &2010 Oct 31.57& & & 46304& & 0.03 $\pm$ 0.01& 2.29 $\pm$ 0.79& $<$ 0.07 & 1.060& 0.141 \\
HCN&SMT & 3-2& 2010 Oct 22.54 & 265886.4& 28.4& 2492 &0.74 &0.24 $\pm$ 0.01 & 1.20 $\pm$ 0.30 & 0.23 $\pm$ 0.03 & 1.062& 0.121 \\
& & & 2010 Oct 29.35& & & 2750& &0.19 $\pm$ 0.05 & 1.06 $\pm$ 0.30& 0.21 $\pm$ 0.01& 1.059& 0.134\\
& && 2010 Nov 1.32& & & 2955& &0.14 $\pm$ 0.03& 1.17 $\pm$ 0.10& 0.16 $\pm$ 0.03& 1.060& 0.144\\
& & &2010 Nov 1.68& & & 2976& & 0.30 $\pm$ 0.04& 1.21 $\pm$ 0.10& 0.36 $\pm$ 0.04& 1.060& 0.144\\
& & &2010 Nov 4.65& & & 3212& & 0.37 $\pm$ 0.01& 1.40 $\pm$ 0.30& 0.38 $\pm$ 0.03& 1.063& 0.156\\
$o$-$\rm H_2CO$ & SMT&$\rm 3_{1,2}-2_{1,1}$& 2010 Oct 29.55& 225697.8& 33.3& 3264& &0.01 $\pm$ 0.03& 1.96 $\pm$ 0.30&0.03 $\pm$ 0.01 & 1.059& 0.134\\
HCN&12m& 1-0&2010 Oct 22.31& 88631.8& 70.7& 6208& 0.95& 0.05 $\pm$ 0.01& 1.50 $\pm$ 0.30 &0.07 $\pm$ 0.01 & 1.062& 0.121\\
&& & 2010 Oct 23.33& & & 6260& & 0.08 $\pm$ 0.02& 0.50 $\pm$ 0.20&0.04 $\pm$ 0.01 & 1.061& 0.122\\
&& & 2010 Oct 24.31& & & 6311& & 0.07 $\pm$ 0.02& 0.90 $\pm$ 0.20 & 0.07 $\pm$ 0.01& 1.060& 0.123\\
&& & 2010 Oct 28.3& & & 6732& & 0.05 $\pm$ 0.03& 1.62 $\pm$ 0.30& 0.08 $\pm$ 0.05& 1.059& 0.131\\
HCN&12m& 2-1& 2010 Oct 23.54& 177261.2& 35.4& 3130& 0.65& 0.35 $\pm$ 0.02& 1.00 $\pm$ 0.40& 0.34 $\pm$ 0.07& 1.061& 0.122\\
CS& 12m& 2-1& 2010 Oct 22.53& 97980.9& 64.0& 5616& 0.88 & 0.07 $\pm$ 0.02& 0.40 $\pm$ 0.20& 0.03 $\pm$ 0.01 &1.062& 0.121\\
HCN& JCMT& 4-3& 2010 Oct 21.48& 354505.5 & 14.1& 1242& 0.63& 0.61 $\pm$ 0.03& 1.16 $\pm$ 0.06 &0.75 $\pm$ 0.05& 1.063& 0.121\\
& & & 2010 Oct 22.43& & & 1242& & 0.59 $\pm$ 0.06& 1.20 $\pm$ 0.20& 0.79 $\pm$ 0.13&1.062& 0.121\\
& & & 2010 Oct 22.78& & & 1252& & 0.97 $\pm$ 0.08 & 1.30 $\pm$ 0.20& 1.31 $\pm$ 0.17&1.061& 0.122\\
& & & 2010 Oct 23.41& & & 1252& & 0.57 $\pm$ 0.04& 1.20 $\pm$ 0.10& 0.73 $\pm$ 0.07&1.061& 0.122\\
& & & 2010 Oct 23.77& & & 1262& & 0.55 $\pm$ 0.04 & 1.20 $\pm$ 0.10& 0.67 $\pm$ 0.08&1.061& 0.123\\
& & & 2010 Nov 2.44& & & 1519& & 0.79 $\pm$ 0.05& 1.20 $\pm$ 0.10& 1.03 $\pm$ 0.10&1.061& 0.148\\
& & & 2010 Nov 2.77& & & 1529& & 0.37 $\pm$ 0.05& 1.40 $\pm$0.20& 0.55 $\pm$ 0.11&1.061& 0.149\\
& & & 2010 Nov 3.62& & & 1560& & 0.30 $\pm$ 0.04 & 1.10 $\pm$ 0.20& 0.33 $\pm$ 0.07&1.062& 0.152\\
& & & 2010 Nov 4.44& & & 1601& & 0.48 $\pm$ 0.05 & 1.30 $\pm$ 0.20& 0.67 $\pm$ 0.11&1.063& 0.156\\
& & 3-2& 2010 Oct 25.56& 265886.4& 18.9& 1716&0.69 & 0.33 $\pm$ 0.05 & 1.20 $\pm$ 0.20& 0.42 $\pm$ 0.09&1.059& 0.125\\
$p$-$\rm H_2CO$ & JCMT&$\rm 5_{0,5}-4_{0,4}$ & 2010 Oct 21.56& 362736.0& 13.8& 1214& & 0.31 $\pm$ 0.01& 0.03 $\pm$ 0.50&0.008 $\pm$ 0.002 & 1.063& 0.121\\
& && 2010 Oct 23.57& & 13.8& 1234& & 0.04 $\pm$ 0.01& 0.39 $\pm$ 0.12 & 0.016 $\pm$ 0.007&1.061& 0.123\\
& && 2010 Nov 2.69& & 13.8& 1494& & 0.05 $\pm$ 0.08& 1.00 $\pm$ 0.50 & 0.049 $\pm$ 0.008&1.061& 0.149\\
$o$-$\rm H_2CO$ & JCMT&$\rm 5_{1,5}-4_{1,4}$& 2010 Nov 2.58& 351768.6& 14.2& 1531& & 0.05 $\pm$ 0.01& 1.60 $\pm$ 0.30& 0.079 $\pm$ 0.018& 1.061& 0.148\\
DCN& JCMT& 5-4& 2010 Oct 22.58& 362046.5& 13.8& 1216& & -& -&$<$ 0.01 & 1.062& 0.121\\
& & & 2010 Oct 23.57& & 13.8& 1236& & -& - & $<$ 0.01&1.061& 0.123\\
& & & 2010 Nov 2.69& & 13.8& 1497& & -& - & $<$ 0.03&1.061& 0.149\\
& & & 2010 Nov 4.67& & 13.8& 1568& & -& - & $<$ 0.04&1.063& 0.156\\
HNC& JCMT& 4-3& 2010 Nov 2.69& 362630.3& 13.8& 1495& & -& - &$<$ 0.03 &1.061& 0.149\\
& & & 2010 Nov 4.67& & 13.8& 1565& & -& -& $<$ 0.04&1.063& 0.156\\
\enddata
\end{deluxetable}

\section{Analysis}
\subsection{Column densities, Production rates and Abundances}

The production rate for water was indirectly measured from the observation of its photodissociation product OH with the GBT. We used the equation of \cite{Tacconi-Garman_1990} to determine the total OH production rate ($Q_{OH}$) for each observation:

\begin{equation}
\scriptsize
Q_{OH} =\left(\frac{7.06 \times 10^{25}}{i} \right) \left( \frac{3.3}{T_{BG}} \right) \left( \frac{10^5} {\tau_{OH}} \right) \left( \frac{\Delta}{1}\right)^2 \left(\frac{I_{OH}}{1} \right) 
\label{Ntot}
\end{equation}
where $i$ is the inversion of the ground state, $T_{BG}$ (K) is the background temperature, $\tau_{OH}$ (s) is the lifetime of the OH molecule, $\Delta$ (AU) is the Earth-comet distance and $I_{OH}$ (mJy km $\rm s^{-1}$) is the integrated flux. The dominant source of error in this formalism comes from the accuracy of the integrated flux measurement, as well as the inversion parameter $i$, although the latter is negligible for these data.

Given that the values of the heliocentric velocity ($R_{dot}$) of the comet ranged from -5 to +2 (Table \ref{OH_Results}), a polynomial function was fit to the data presented in Table 5 of \cite{Schleicher_1988} between these values in order to find the inversion at the comet's heliocentric velocity on the date and UT time of the observations. This inversion value is given in Table \ref{OH_Results}.The OH photodissociation lifetime ($\tau_{OH}$) is 1.2$\times$ $10^5$ $R_h^2$ s (taken from \citealt{Tacconi-Garman_1990}), and is used in the final production rate calculation, and we assume $T_{BG}$ $=$ 2.7 K. The measured intensity and line width of each transition is given in Table \ref{Obs_parameters}. The $Q_{OH}$ value is then corrected by a factor $P$, which is defined as the fraction of molecules in the coma that are contained within the GBT beam. We computed $P$ using a Monte-Carlo Haser model \citep{Haser_1957} for daughters molecules, assuming $\tau_{H_2O}$ = 8.0 $\times$ $10^4$ $R_h^2$ (s) \citep{Huebner_1992}, $v_{OH}$ = 0.95 km $s^{-1}$ and $v_{H_2O}$ = 0.80 km $s^{-1}$ \citep{Crovisier_2013}. The final production rates for OH on each date are given in Table \ref{OH_Results}. $Q_{OH}$ shows large diversity with time. The average measured production rate of OH over this monitoring campaign is $Q_{OH}$ = (7.03 $\pm$ 0.44) $\times$ $\rm 10^{27}$ $\rm s^{-1}$. The conversion to $Q_{H_2O}$ is made by dividing the $Q_{OH}$ by the branching ratio of water dissociation to OH. The photodissociation of the $\rm H_2O$ molecule into H and OH is the most important process accounting for 85.5$\%$ of all water molecules that are dissociated \citep{Harris_2002}. The mean water production rate was therefore $Q_{H_2O}$ = (8.22 $\pm$ 0.51) $\times$ $\rm 10^{27}$ $\rm s^{-1}$.

\begin{table*}
\caption{OH and $\rm H_2O$ production rates with the GBT toward comet 103P/Hartley 2. } 
\label{OH_Results}
\centering
\scriptsize
\begin{tabular}{llllllll}
\tableline\tableline
Line &Telescope & UT Date & $R_{dot}$&$i$ calculated & P&$Q_{OH}$ &$ Q_{H_2O}$ \\
& &  &(km $\rm s^{-1}$) && & ($\rm s^{-1}$) &($\rm s^{-1}$) \\
\hline
OH & GBT& 2010 Oct 13.04 & -4.99 & -0.26 & 0.063 &(6.99 $\pm$ 0.73) $\times$ $\rm 10^{27}$ & (8.17 $\pm$ 0.85) $\times$ $\rm 10^{27}$ \\
& &2010 Oct 15.04 &-4.11 & -0.23& 0.061&(9.13 $\pm$ 0.91) $\times$ $\rm 10^{27}$ & (1.07 $\pm$ 0.11) $\times$ $\rm 10^{28}$ \\
& &2010 Oct 16.40 & -3.70 & -0.21 & 0.059&(6.26 $\pm$ 0.81) $\times$ $\rm 10^{27}$ & (7.32 $\pm$ 0.95) $\times$ $\rm 10^{27}$ \\
& &2010 Oct 19.44 & -2.89 &-0.15 & 0.058&(5.73 $\pm$ 1.07) $\times$ $\rm 10^{27}$ & (6.71 $\pm$ 1.25) $\times$ $\rm 10^{27}$ \\

\hline

\end{tabular}

\end{table*}

\begin{figure*}
\centering
\includegraphics[scale=0.5]{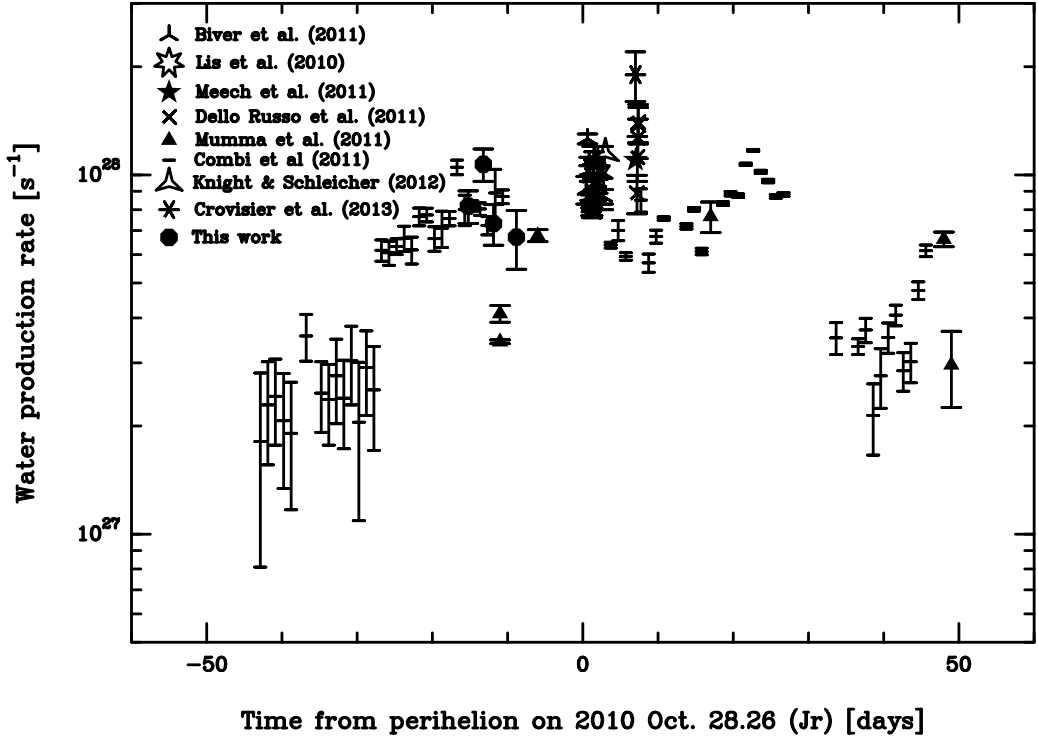}
\caption{Water production rates measured in comet 103P/Harley 2 as a function of the time from perihelion on 2010 Oct. 28.26 (defined as Jr). Between Jr+1 and Jr+3 days, the observations are essentially from \cite{Biver_2011} who studied the short-term variability of water. \cite{Lis_2010} and \cite{Knight_2012}, derived $Q_{H_2O}$ = (1.0 $\pm$ 0.2) $\times$ $10^{28}$ $\rm s^{-1}$ and at Jr+3 and $Q_{H_2O}$ = 1.15 $\times$ $10^{28}$ $\rm s^{-1}$ at Jr+2 days, respectively. At Jr+7 days, the observations are esentially from \cite{Dello_2011}. \cite{Meech_2011} and \cite{Crovisier_2013} derived $Q_{H_2O}$ = 1.2 $\times$ $10^{28}$ $\rm s^{-1}$ and $Q_{H_2O}$ = (1.9 $\pm$ 0.3) $\times$ $10^{28}$ $\rm s^{-1}$, respectively.}
\label{Qwater}
\end{figure*}

Water production was directly measured in this comet from submillimeter rotational transitions from space with Odin \citep{Biver_2011}, with the three instruments of Herschel \citep{Lis_2010, Hartogh_2011, Meech_2011}, and from rovibrational lines in the infrared from the ground 
\citep{Dello_2011, Mumma_2011}. It was indirectly measured from the observation of its photodissociation products: of OH in the near-UV \citep{Knight_2012}, of H from the $\rm Ly$ --$ \alpha$ line observed by SOHO/SWAN \citep{Combi_2011} and of OH at 18 cm \citep[here and][]{Crovisier_2013}. Published water production rates from concurrent observations within $\pm$50 days of perihelion (labeled as Jr) ranged from 0.1 to 1.9 $\times$ $10^{28}$ $\rm s^{-1}$ and are summarized in Figure \ref{Qwater}. The values from this work are consistent with the value derived by \cite{Mumma_2011} and \cite{Combi_2011}. The observed variation in $Q_{H_2O}$ as been attributed to the rotation of the nucleus \citep{Mumma_2011, Biver_2011}.\\




Beam-averaged column densities were derived for HCN, $\rm H_2CO$, CS, DCN and HNC assuming that the cometary coma filled the beams of the respective telescopes. In the case of the upper limits, a 3$\sigma$ rms was assumed to derive the integrated intensity of DCN and HNC, set at the linewidth of $\approx$ 1 km $s^{-1}$. The column density for the observations was calculated from:
\begin{equation}
\scriptsize
N_{tot} =\frac{8 \pi k \nu^2 \int T_R \Delta v_{1/2} \zeta_{rot}}{ h c^3 A_{ul} g_{up} \eta_c} e^{E_{up}/k T_{rot}},
\label{Ntot}
\end{equation}
where $k$ is the Boltzmann constant, $\zeta_{rot}$ is the partition function, $h$ is the Planck constant, $c$ is the speed of the light, $A_{ul}$ [$\rm s^{-1}$] is the Einstein coefficient, $g_{up}$ is the statistical weight, $T_{rot}$ [K] is the rotational excitation temperature, $E_{up}$ [$\rm cm^{-1}$] is the upper state energy, and $N_{tot}$ [$\rm cm^{-2}$] is the total number of molecules observed in the beam. \cite{Drahus_2012} concluded that the rotational temperature from $\rm CH_3OH$ (157.225 GHz; IRAM 30m telescope) varied strongly, due to the nucleus rotation, with the average value being 47~K (between 19-179~K). Additionally, \cite{Boissier_2014} derived $T_{rot}$ from $\rm CH_3OH$ (157.225 GHz; IRAM 30m telescope and the Plateau de Bure) as a function of the radii of the projected beam sizes on the coma. They concluded that the increase of $T_{rot}$ from $\approx$ 35K to $\approx$ 46~K is due to the increase of the beam radii from $\approx$ 150~km to $\approx$ 1500~km. In this analysis, we assumed $T_{rot}$ = 50~K, which is the average from $\rm CH_3OH$ observations conducted at the same facilities during our observations (Chuang et al. 2014, in preparation). This is consistent with other results considering a diameter of the projected beam sizes on the coma between $\approx$ 1200~km to $\approx$ 46,400~km (Table \ref{Obs_parameters}). The column densities derived for HCN, $\rm H_2CO$, CS, DCN and HNC are listed in Table \ref{Qgas}. The minimum ($T_{rot}$ = 19K) and maximum rotational ($T_{rot}$ = 179~K) temperatures obtained by \cite{Drahus_2012} provide $ N_{tot}$ = (6.29 $\pm$ 0.34) $\times$ $\rm 10^{11}$ $\rm cm^{-2}$ and $ N_{tot}$ = (2.32 $\pm$ 0.12) $\times$ $\rm 10^{11}$ $\rm cm^{-2}$, respectively on 2010 October 22 for HCN: J $=3-2$. The D/H and $ortho$:$para$ ratio are discussed in sections \ref{SECTION_DHR} and \ref{SECTION_OPR}, respectively.

Most of the molecular species measured are considered to be parent species, thus production rates were determined from a Monte Carlo model \citep{Milam_2004, Milam_2006} that traces the trajectories of molecules within the telescope beam, ejected from the comet nucleus. The observed column density is then matched for an output molecular production rate, $Q$. $\rm H_2CO$ is considered as both a parent and extended source species, so production rates were derived for both cases employing the models of \cite{Milam_2006}.
Table \ref{Qgas} summarizes the production rates of the molecules observed toward comet 103P, the production rate of water and the ratio $ Q/Q_{H_2O}$ with the SMT, 12m, and JCMT. This model assumes isotropic outgassing, which is reasonable for the analysis of these data since the model simulates the observed column densities within a large beam (with respect to the comet). The uncertainties introduced into the modeled production rates are dominated by the errors on the measured column densities.

\begin{deluxetable}{lllllllll}
\tabletypesize{\scriptsize}
\rotate
\tablecaption{Column densities, photodissociation rates, production rates for HCN, $\rm H_2CO$, CS, DCN and HNC, production rate of water and the ratio $ Q/Q_{H_2O}$ with the SMT, 12m, and JCMT. $Q_{H_2O}$ are taken from comparable date (Figure \ref{Qwater}). The photodissociation rates are from Photo Ionization/Dissociation rates. $*$ denotes the EPOXI flyby.\label{Qgas}}
\tablewidth{0pt}
\tablehead{
\colhead{Line} & \colhead{Telescope} & \colhead{ Transition } & \colhead{UT Date } & \colhead{$N_{tot}$} &
\colhead{Rate} & \colhead{Q} & \colhead{$ Q/Q_{H_2O}$}  \\

\colhead{} & \colhead{} & \colhead{} & \colhead{ } & \colhead{($cm^{-2}$) } &
\colhead{$\rm (s^{-1})$} & \colhead{$\rm (s^{-1})$} & \colhead{$(\%$) } 
}
\startdata
HCN &SMT& 3-2& 2010 Oct 22.54& (2.72 $\pm$ 0.38) $\times$ $\rm 10^{11}$& 1.31 $\times$ $\rm 10^{-05}$& (4.25 $\pm$ 0.59) $\times$ $\rm 10^{24}$ & (6.27 $\pm$ 0.91) $\times$ $\rm 10^{-02}$\\
& & &2010 Oct 29.35& (2.45 $\pm$ 0.13) $\times$ $\rm 10^{11}$ & & (3.77 $\pm$ 0.20) $\times$ $\rm 10^{24}$ & (5.91 $\pm$ 0.34) $\times$ $\rm 10^{-02}$\\
& && 2010 Nov 1.32& (1.93 $\pm$ 0.36) $\times$ $\rm 10^{11}$ & & (3.55 $\pm$ 0.66) $\times$ $\rm 10^{24}$ & (5.06 $\pm$ 0.99) $\times$ $\rm 10^{-02}$ \\
& & & 2010 Nov 1.68& (4.34 $\pm$ 0.48) $\times$ $\rm 10^{11}$ && (8.18 $\pm$ 0.90) $\times$ $\rm 10^{24}$ & (1.17 $\pm 0.15)$ $\times$ $\rm 10^{-01}$\\
& & & 2010 Nov 4.65& (4.49 $\pm$ 0.40) $\times$ $\rm 10^{11}$ && (9.94 $\pm$ 0.88) $\times$ $\rm 10^{24}$ & (7.83 $\pm 1.20)$ $\times$ $\rm 10^{-02}$\\
$o$-$\rm H_2CO$ & SMT&$\rm 3_{1,2}-2_{1,1}$& 2010 Oct 29.55& (2.39 $\pm$ 0.35) $\times$ $\rm 10^{11}$ &2.15 $\times$ $\rm 10^{-04}$& (1.07 $\pm$ 0.16) $\times$ $\rm 10^{25}$ & (1.68 $\pm$ 0.25) $\times$ $\rm 10^{-01}$\\
HCN& 12m& 1-0 &2010 Oct 22.31 & (3.94 $\pm$ 0.67) $\times$ $\rm 10^{11}$& 1.31 $\times$ $\rm 10^{-05}$& (1.97 $\pm$ 0.34) $\times$ $\rm 10^{25}$ & (2.91 $\pm 0.51)$ $\times$ $\rm 10^{-01}$\\
& & &2010 Oct 23.33& (2.33 $\pm$ 0.78) $\times$ $\rm 10^{11}$& & (4.38 $\pm$ 1.45) $\times$ $\rm 10^{24}$ & (6.45 $\pm$ 2.17) $\times$ $\rm 10^{-02}$\\
& && 2010 Oct 24.31& (3.68 $\pm$ 0.67) $\times$ $\rm 10^{11}$& &(1.17 $\pm$ 0.22) $\times$ $\rm 10^{25}$ & (1.73 $\pm$ 0.32) $\times$ $\rm 10^{-01}$ \\
& & & 2010 Oct 28.3& $<$ 4.13 $\times$ $\rm 10^{11}$& & $<$ 2.43 $\times$ $\rm 10^{25}$ & $<$ 3.81 $\times$ $\rm 10^{-01}$\\
HCN& 12m& 2-1& 2010 Oct 23.54& (7.95 $\pm$ 1.75) $\times$ $\rm 10^{11}$& & (1.32 $\pm$ 0.29) $\times$ $\rm 10^{25}$ & (1.95 $\pm$ 0.44) $\times$ $\rm 10^{-01}$\\
CS& 12m& 2-1& 2010 Oct 22.53& (3.35 $\pm$ 1.10) $\times$ $\rm 10^{11}$& 1.00 $\times$ $\rm 10^{-04}$& (9.32 $\pm$ 3.10) $\times$ $\rm 10^{24}$ & (1.37 $\pm$ 0.45) $\times$ $\rm 10^{-01}$ \\
HCN &JCMT& 4-3& 2010 Oct 21.48& (8.28 $\pm$ 0.55) $\times$ $\rm 10^{11}$& 1.31 $\times$ $\rm 10^{-05}$& (6.23 $\pm$ 0.42) $\times$ $\rm 10^{24}$ & (9.19 $\pm$ 0.72) $\times$ $\rm 10^{-02}$\\
&& & 2010 Oct 22.43& (8.73 $\pm$ 1.44) $\times$ $\rm 10^{11}$& & (6.80 $\pm$ 1.21) $\times$ $\rm 10^{24}$ & (1.01 $\pm$ 0.18) $\times$ $\rm 10^{-01}$\\
& & & 2010 Oct 22.78& (1.45 $\pm$ 0.19) $\times$ $\rm 10^{12}$& & (1.23 $\pm$ 0.16) $\times$ $\rm 10^{25}$ & (1.81 $\pm$ 0.25) $\times$ $\rm 10^{-01}$ \\
& & & 2010 Oct 23.41& (7.40 $\pm$ 0.88) $\times$ $\rm 10^{11}$& & (5.81 $\pm$ 0.69) $\times$ $\rm 10^{24}$ & (8.75 $\pm$ 1.08) $\times$ $\rm 10^{-02}$\\
& & & 2010 Oct 23.77& (1.14 $\pm$ 0.10) $\times$ $\rm 10^{12}$& & (9.02 $\pm$ 0.91) $\times$ $\rm 10^{24}$ & (1.33 $\pm$ 0.14) $\times$ $\rm 10^{-01}$\\
& & & 2010 Nov 2.44& (1.14 $\pm$ 0.11) $\times$ $\rm 10^{12}$ && (1.09 $\pm$ 0.11) $\times$ $\rm 10^{25}$ & (1.55 $\pm$ 0.15) $\times$ $\rm 10^{-01}$\\
& & & 2010 Nov 2.77& (6.07 $\pm$ 1.21) $\times$ $\rm 10^{11}$& & (6.79 $\pm$ 1.35) $\times$ $\rm 10^{24}$ & (1.15 $\pm$ 0.23) $\times$ $\rm 10^{-01}$\\
& & & 2010 Nov 3.62& (3.64 $\pm$ 0.77) $\times$ $\rm 10^{11}$& & (3.29 $\pm$ 0.70) $\times$ $\rm 10^{24}$ & (3.72 $\pm$ 0.91) $\times$ $\rm 10^{-02}$ \\
& & & 2010 Nov 4.44$^*$& (7.40 $\pm$ 1.21) $\times$ $\rm 10^{11}$& & (8.07 $\pm$ 1.32) $\times$ $\rm 10^{24}$ & (9.13 $\pm$ 1.87) $\times$ $\rm 10^{-02}$\\

&& 3-2& 2010 Oct 25.56& (5.36$\pm$ 1.15) $\times$ $\rm 10^{11}$& & (5.78 $\pm$ 1.24) $\times$ $\rm 10^{24}$ & (8.53 $\pm$ 1.86) $\times$ $\rm 10^{-02}$\\

$p$-$\rm H_2CO$ & JCMT&$\rm 5_{0,5}-4_{0,4}$& 2010 Oct 21.56& (1.14 $\pm$ 0.26) $\times$ $\rm 10^{11}$& 2.15 $\times$ $\rm 10^{-04}$& (2.65 $\pm$ 0.60) $\times$ $\rm 10^{23}$ & (3.91 $\pm$ 0.89) $\times$ $\rm 10^{-03}$\\
& JCMT&& 2010 Oct 23.57& (2.41 $\pm$ 1.05) $\times$ $\rm 10^{11}$ & & (1.02 $\pm$ 0.43) $\times$ $\rm 10^{24}$ & (1.50 $\pm$ 0.53) $\times$ $\rm 10^{-02}$ \\
& JCMT&& 2010 Nov 2.69& (7.38 $\pm$ 1.21) $\times$ $\rm 10^{11}$& & (7.54 $\pm$ 1.23) $\times$ $\rm 10^{24}$ & (1.08 $\pm$ 0.19) $\times$ $\rm 10^{-01}$\\
$o$-$\rm H_2CO$ & JCMT&$\rm 5_{1,5}-4_{1,4}$& 2010 Nov 2.58& (5.21 $\pm$ 1.19) $\times$ $\rm 10^{11}$ & & (7.99$\pm$ 1.82) $\times$ $\rm 10^{24}$ & (1.14 $\pm$ 0.27) $\times$ $\rm 10^{-02}$\\
DCN &JCMT& 5-4& 2010 Oct 22.58& $<$ 1.53 $\times$ $\rm 10^{10}$ & 1.31 $\times$ $\rm 10^{-05}$& $<$ 1.36 $\times$ $\rm 10^{23}$& $<$ 2.01 $\times$ $\rm 10^{-03}$\\
& & & 2010 Oct 23.57& $<$ 1.53 $\times$ $\rm 10^{10}$& & $<$ 1.38 $\times$ $\rm 10^{23}$ & $<$ 2.04 $\times$ $\rm 10^{-03}$ \\
& & & 2010 Nov 2.69& $<$ 3.44 $\times$ $\rm 10^{10}$& & $<$ 3.77 $\times$ $\rm 10^{23}$& $<$ 5.38 $\times$ $\rm 10^{-03}$ \\
& & & 2010 Nov 4.67$^*$& $<$ 4.59 $\times$ $\rm 10^{10}$& & $<$ 5.27 $\times$ $\rm 10^{23}$& $<$ 4.15 $\times$ $\rm 10^{-03}$\\
HNC& JCMT& 4-3& 2010 Nov 2.69& $<$ 3.09 $\times$ $\rm 10^{10}$ &1.31 $\times$ $\rm 10^{-05}$& $<$ 3.38 $\times$ $\rm 10^{23}$ & $<$ 4.82 $\times$ $\rm 10^{-03}$\\
& & & 2010 Nov 4.67$^*$& $<$ 4.33 $\times$ $\rm 10^{10}$& & $<$ 5.08 $\times$ $\rm 10^{23}$& $<$ 4.00 $\times$ $\rm 10^{-03}$
\enddata
\end{deluxetable}



\subsection{103P/Hartley 2 previous measurements}
\label{103P/Hartley 2 measurements}

\cite{Crovisier_2013} observed the short-term variation of the OH production rate at Nancay in 2010 October. The production rate increased steeply and progressively before perihelion, reaching a maximum just before the EPOXI flyby. The water production rate preceding perihelion measured by SOHO/SWAN also shows an increase by a similar amount \citep{Combi_2011}. By using the GBT data, one can also see that $Q_{OH}$ increased as the comet was approaching the Sun (Appendix \ref{Appendix_OH}). The time step of the GBT data cannot be used to correlate the temporal evolution of the water production rate with the rotational period of the comet, as studied with some other data, due to the long lifetime of OH and the large beam telescope beam. For example, \cite{Mumma_2011} reproduced the temporal evolution of water with a cometary rotational period of 18 hours.

\cite{Mumma_2011} also showed that the production rates for ethane, HCN, and methanol vary in a manner consistent with independent measures of nucleus rotation. Similar to OH, the time steps of our HCN observations were not adequate to study the short-term variability of HCN. \cite{Drahus_2012} and \cite{Boissier_2014} studied in detail the short-term variability of HCN. \cite{Drahus_2012} observed a perfect phasing between the $\rm CH_3OH$ and HCN production rates. \cite{Boissier_2014} compared the production curves of HCN, $\rm H_2O$, $\rm CH_3OH$, amd $ \rm CO_2$. They showed that the curves of $\rm H_2O$, $\rm CH_3OH$ and HCN are in phase, but there is a delay of $\approx$ 1.7 hrs with $ \rm CO_2$. They concluded that this delay can be due to the production of $\rm H_2O$, $\rm CH_3OH$, and HCN from subliming icy grains, whereas $\rm CO_2$ molecules were released from the nucleus. By studying the spatial distribution of parent volatiles in comet 103P observed in the infrared, \cite{Kawakita_2013} reported that the spatial distributions of HCN, $\rm C_2H_2$, $\rm C_2H_6$ were extended in the anti-solar direction, while the spatial distributions of $\rm H_2O$ and $\rm CH_3OH$ were significantly extended in the solar direction. They also concluded that the molecules are produced from the sublimation of icy grains but that there might be two distinct phases of ice in 103P; one enriched in $\rm H_2O$ and $\rm CH_3OH$, and another enriched in more volatile species (HCN, $\rm C_2H_2$, $\rm C_2H_6$). This asymmetry in the line profile mentioned by \cite{Kawakita_2013} can be observed for HCN with the JCMT (Appendix \ref{Appendix_HCN}.).

In Table \ref{Qgas} we can see that the abundance ratio of HCN with water shows large variation with time, so the average is more representative of the global chemistry. This range can be explained by the asymmetric spatial profiles of the molecules \citep{Kawakita_2013}, because the molecules are from different parts of the nucleus, or because of the sublimation of icy grains. The significant contribution of grain sublimation to the production of volatiles is supported by numerous measurements \citep{Ahearn_2011, Mumma_2011, Dello_2011, Drahus_2012, Knight_2012}. By averaging the abundance ratio of HCN and $o-$$\rm H_2CO$ with water, we obtain Q(HCN)/Q($\rm H_2O$) = 0.13 $\pm$ 0.03 $\%$ and Q($\rm H_2CO$)/Q($\rm H_2O$) = 0.14 $\pm$ 0.03 $\%$. \cite{Boissier_2014} derived an average abundance of HCN relative to water from millimeter observations consistent with our results. They detected HCN on 2010 October 23, November 4 and 5 with the IRAM interferometer located at the Plateau de Bure and obtained Q(HCN)/Q($\rm H_2O$) = 0.16 $\%$. The HCN abundances from millimeter observations were lower than the values from infrared data (Q(HCN)/Q($\rm H_2O$) $\approx$ 0.3 $\%$ \citealt{Dello_2011, Dello_Russo_2013, Mumma_2011, Kawakita_2013}), by the typical factor of $\approx2$ \citep{Villanueva_2013}. We deduced an average abundance of $\rm H_2CO$ relative to water in good agreement with \cite{Dello_Russo_2013} and \cite{Kawakita_2013} $\approx$ 0.11$\%$, but the value from this work is lower than the one from \cite{Mumma_2011}. The discrepancy between radio and IR results is not surprising given the strong temporal variability of the coma and differences in field of view. However, by comparison with other comets, we find the same approximate relationships between species in 103P as observed in the infrared: $\rm H_2CO$ depleted and HCN normal. Abundance ratios for other primary volatiles were obtained by \cite{Dello_2011} and \cite{Kawakita_2013}; they concluded that comet 103P is $\rm C_2H_2$ normal, $\rm C_2H_6$ normal, $\rm NH_3$ normal, $\rm CH_3OH$ normal, $\rm CH_4$ depleted and CO depleted.


\subsection{D/H ratio}
\label{SECTION_DHR}
We used the JCMT data to compute an upper limit for the D/H ratio in HCN and DCN. Both species were sampled the same day. We average the gas production rate of DCN on UT 2010 October 22.58 and UT 2010 October 23.57 (Figure in Appendix \ref{Appendix_DCN}). The D/H ratios from HCN are given in Table \ref{D-H}. As discussed in section \ref{103P/Hartley 2 measurements}, we note variability of the HCN with time, and this short-term variabilty has been studied in detail by \cite{Boissier_2014} and \cite{Drahus_2012}. We measured only upper limits for DCN, for which we assumed a constant HCN production rate over the period between the HCN and DCN observations on a given day.

\begin{table*}
\centering
\caption{D/H ratio in comet 103P/Harley 2. We average $\rm Q_{DCN}$ on UT 2010 October 22.58-23.57. } \label{D-H}
\scriptsize
\begin{tabular}{lllll}
\tableline\tableline
UT Date & $\rm Q_{HCN}$ & UT Date & $\rm Q_{DCN}$ & D/H \\
& ($\rm s^{-1}$) & & ($\rm s^{-1}$) & \\
\hline
2010 Oct 22.43 & 6.80 $\times$ $\rm 10^{24}$ &2010 Oct 22.58-23.57 & $<$1.37 $\times$ $\rm 10^{23}$ &$<$2.01 $\times$ $\rm 10^{-02}$\\
2010 Oct 22.78 & 1.23 $\times$ $\rm 10^{25}$ &&$<$1.37 $\times$ $\rm 10^{23}$ &$<$1.11 $\times$ $\rm 10^{-02}$\\
2010 Oct 23.41 & 5.81 $\times$ $\rm 10^{24}$ & &$<$1.37 $\times$ $\rm 10^{23}$ &$<$2.36 $\times$ $\rm 10^{-02}$\\
2010 Oct 23.77 & 9.02 $\times$ $\rm 10^{24}$ &&$<$1.37 $\times$ $\rm 10^{23}$ &$<$1.52 $\times$ $\rm 10^{-02}$\\
2010 Nov 2.44 & 1.09 $\times$ $\rm 10^{25}$ &2010 Nov 2.69&$<$3.77 $\times$ $\rm 10^{23}$ &$<$3.46 $\times$ $\rm 10^{-02}$\\
2010 Nov 2.77 & 6.79 $\times$ $\rm 10^{24}$ &2010 Nov 2.69&$<$3.77 $\times$ $\rm 10^{23}$ &$<$5.55 $\times$ $\rm 10^{-02}$\\
2010 Nov 4.44 & 8.07 $\times$ $\rm 10^{24}$ &2010 Nov 4.67& $<$5.27 $\times$ $\rm 10^{23}$& $<$6.63 $\times$ $\rm 10^{-02}$\\
\hline
\end{tabular}

\end{table*}

Relevant deuterium fractionation ratios have been measured for both comets and the interstellar medium. \cite{Crovisier_2004a}, \cite{Kawakita_2005}, \cite{Kawakita_2009}, \cite{Gibb_2012} and \cite{Bonev_2009} have presented upper limits for the production rates of several cometary molecules. D/H isotope ratios have been measured for two cometary molecules in 103P: water and hydrogen cyanide. Compared to models of interstellar chemistry, both the measured HDO/$\rm H_2O$ and DCN/HCN ratios are compatible with ion-molecule chemistry in gas at about 30-35 K. With this work we measured an upper limit for the DCN/HCN ratio of $<$0.01. Our value is consistent with previous measurements [e.g., D/H = 0.002 in comet Hale-Bopp; \cite{Meier_1998}, and 1.6 $\times$ $\rm 10^{-4}$ in comet 103P; \cite{Hartogh_2011}], although it does not place any new constraints. By comparison, a range of (0.4 - 7.0) $\times$ $10^{-2}$ for the DCN/HCN ratio was determined in the interstellar medium \citep{Roberts_2002, Jorgensen_2004}.

\subsection{OPR ratio from $\rm H_2CO$}
\label{SECTION_OPR}
Molecules that contain multiple H atoms can be spectroscopically distinguished based on the orientation of the spins of their H nuclei, either aligned in parallel or anti-parallel states (e.g. $\rm H_2$, $\rm H_2CO$, $\rm H_2O$), giving rise to $ortho$ and $para$ forms. Quantum-mechanically, transitions between the different forms are strictly forbidden giving rise to two distinct spectra and $ortho:para$ ratios (OPRs), which define the spin temperature ($T_{spin}$) of that species. This means that the relative abundances of nuclear spin isomers are expected to remain invariant over time, and the OPR has therefore been long thought to reveal information about the temperature of formation of cometary ices. However, some processes are hypothesized to change the OPR without breaking molecular bonds \citep[e.g.][]{Limbach_2006}, and phase transitions are effective in bringing the molecules into an equilibrated OPR \cite[e.g.][]{Watanabe_2013}. In the study of water molecules trapped in low temperature ($\sim4$ K) solid Ar matrices, \cite{Sliter_2011} found that upon desorption at T $>$ 260~K, fast $ortho-para$ conversion occurred. In addition, theoretical investigations \citep{Anderson_2008} showed that photodesorption of water ice at 10 K would typically lead to bond breaking (with the subsequent loss of OPR information). Of the processes investigated by \citet{Anderson_2008}, only the `kick-out' mechanism would preserve the OPR during desorption at low temperatures. On the other hand, it is not straightforward to extrapolate the high-temperature desorption and Ar matrix-ice studies by \cite{Sliter_2011} to the release of volatiles from cometary ices, and which process dominates the desorption of ices at low temperatures. The cosmogonic significance of OPR and spin temperature in comets is therefore not clear, and the lack of OPR measurements for molecules with smaller $ortho-para$ energy deficiencies (e.g., $\rm H_2CO$ and $\rm CH_3OH$) has further limited our understanding of this indicator in comets.

The OPR of H$_2$CO for Hartley 2 was obtained from consecutive observations of the $ortho$-$\rm H_2CO$ ($\rm 5_{0,5}-4_{0,4}$) and $para$-$\rm H_2CO$ ($\rm 5_{1,5}-4_{1,4}$) lines on November 2 from data obtained at the JCMT (see Table \ref{Obs_parameters}). The periodicity of the comet presents some uncertainty in abundances, though these two observations were in close enough proximity ($<$ 3 hrs). Therefore, the variation in rotation temperature is likely not a major factor. Other observational uncertainties are factored out in the ratio. For these data we obtain an OPR $\approx$ 2.12 $\pm$ 0.59 (1$\sigma$), or $T_{spin}$ $\approx$ 13.5$^{+6.7}_{-3.3}$ K (1$\sigma$), although at only 1.5$\sigma$ our OPR measurement cannot distinguish from equilibrium ($T_{spin}$ $>$ 9 K at 1.5$\sigma$). Similarly, \cite{Villanueva_2012b} have measured $T_{spin}$ $>$ 18 K for $\rm CH_3OH$ in comet C/2001 A2 (LINEAR), which is consistent with that in Hale-Bopp $T_{spin}$ $>$ 15 K \citep{Pardanaud_2007}. Our results are also consistent (within 1$\sigma$) with those obtained for water by \cite{Bonev_2013} of OPR $\approx$ 2.79 $\pm$ 0.13 (mean of five measurements).
 
The complexity of obtaining simultaneous observations of $ortho$ and $para$-$\rm H_2CO$, and the difficulty in obtaining a good signal-to-noise ratio for these data enforces the future need for sensitive, broad-band telescope receivers where these can be readily measured towards any comet. Such measurements will provide the necessary context to better understand the cosmogonic significance (or lack of) of our OPR measurement of $\approx$ 2.12 $\pm$ 0.59 (1$\sigma$).

\section{Comparison among various comets}
Knowledge of the volatile chemistry of Jupiter-Family comets (Kuiper Belt) is very limited from both infrared and radio observations because Jupiter-Family comets have typically lower activity.

\cite{Crovisier_2009} show the compositional diversity among Jupiter-Family comets from radio observations. Before the observation of 103P, about a dozen Jupiter-Family comets have been observed by radio techniques. By comparing the relative abundance of HCN among Jupiter-Family comets, \cite{Crovisier_2009} reported that 2P/Encke, 9P/Tempel 1 and 22P/Kopff are organics-normal, like 103P, while 19P/Borrelly, 21P/Giacobini-Zinner and 73P/Schwassmann-Wachmann 3 are organics-depleted. This suggests distinct processing histories for organics-depleted, organics-normal and organics-enriched comets. \cite{Dello_2011} and \cite{Kawakita_2013} compared the abundances ratios among Jupiter-family sampled from infrared observations. The limited sample of few Jupiter-Family comets shows again that diversity is notable. For example, \cite{Dello_2011} concluded than 73P/Schassmann-Wachmann 3 and 103P are extremely different in their chemical composition.

Correlations between short-term temporal variations in production rates of primary volatiles with nucleus rotation have been suggested only for two Oort cloud comets. \cite{Biver_2009} reported a periodic variation of 40$\%$ in the water production rate in C/2001 Q4 (NEAT), with a period of 19.58 $\pm$ 0.1 hrs. \cite{Anderson_2010} observed a variation in production rates for $\rm H_2O$, CO, $\rm H_2CO$, and $\rm CH_3OH$ in comet C/2002 T7 (LINEAR) with a period of 2.32 days. Comet 103P is the third comet for which periodic variation in production of primary volatiles has been demonstrated, and the first one for which unambiguous association with nucleus rotation can be made through imaging \citep{Ahearn_2011, Harmon_2010, Harmon_2011}.

To date, OPRs have been measured in a few species, namely $\rm NH_3$, $\rm H_2O$, and $\rm CH_4$, for several comets by ground-based observations \citep[see][]{DBM_2004, Dello_Russo_2005, Kawakita_2002, Kawakita_2001, Mumma_2011}. \cite{Dello_Russo_2005} reported from $\rm H_2O$, $T_{spin}$ $>$ 30 K, $T_{spin}$ $=$ $\rm 30^{+15}_{-6}$ K, $T_{spin}$ $=$ $\rm 23^{+4}_{-3}$ K respectively in comets C/1999 H1 (Lee), C/1999 S4 (LINEAR), and C/2001 A2 (LINEAR). \cite{Kawakita_2001} and \cite{Kawakita_2002} derived an OPR of ammonia equal to 1.17 $\pm$ 0.04 and 1.12 $\pm$ 0.03 (or $T_{spin}$ $=$ $\rm 30^{+3}_{-2}$ K) respectively in comet C/1999 S4 (LINEAR) and C/2001 A2 (LINEAR). Some of the OPR radios imply formation temperatures for cometary ice species of 25 $–-$ 35 K, similar to the formation temperature implied by the D/H ratios in $\rm H_2O$ and HCN, but a growing sample of comets also show equilibrated OPR \citep[$T_{spin}$ $>$ 35 K, e.g.][]{Villanueva_2012a, Mumma_1993}. For example, \cite{Mumma_1993} reported a water OPR $=$ 3.2 $\pm$ 0.2 for C/1986 P1 (Wilson) ($T_{spin}$ $>$ 50 K).

\section{Conclusions}

We conducted observations of comet 103P/Hartley 2 at both perihelion and at the time of the EPOXI flyby. We report detections of HCN, $\rm H_2CO$, CS, and OH and upper limits of HNC and DCN, using the ARO Kitt Peak 12m and SMT, JCMT and the GBT towards comet 103P. We derived column densities, production rates, and relative abundances toward comet 103P. We concluded that 103P is normal in HCN and depleted in $\rm H_2CO$ which is in good agreement with other studies. We obtained an upper limit the $\rm (D/H)_{HCN}$ ratio of $<$ 0.01 and an $ortho:para$ ratio of 2.2 $\pm$ 0.59 (1$\sigma$) has been derived from $\rm H_2CO$.

\begin{acknowledgements}
\textit{Acknowledgements}: \\
The National Radio Astronomy Observatory is a facility of the National Science Foundation operated under cooperative agreement by Associated Universities, Inc. The Kitt Peak 12m telescope and the Submillimeter telescope are currently operated by the Arizona Observatory (ARO), Steward Observatory, University of Arizona, with partial funding from the Research Corporation. The James Clerk Maxwell Telescope is operated by the Joint Astronomy Centre on behalf of the Science and Technology Facilities Council of the United Kingdom, the Netherlands Organisation for Scientific Research, and the National Research Council of Canada. This work was supported by the Goddard Center for Astrobiology, by NASA's Planetary Astronomy and Planetary Atmospheres Programs. YJK acknowledges support from NSC grants 100-2119-M-003-001-MY3 and 102-2119-
M-003-008- for this work.

\end{acknowledgements}


\begin{appendix}
\section{Detection of OH}
\label{Appendix_OH}
\begin{figure}[h!]
  \centering
\includegraphics[scale=0.7]{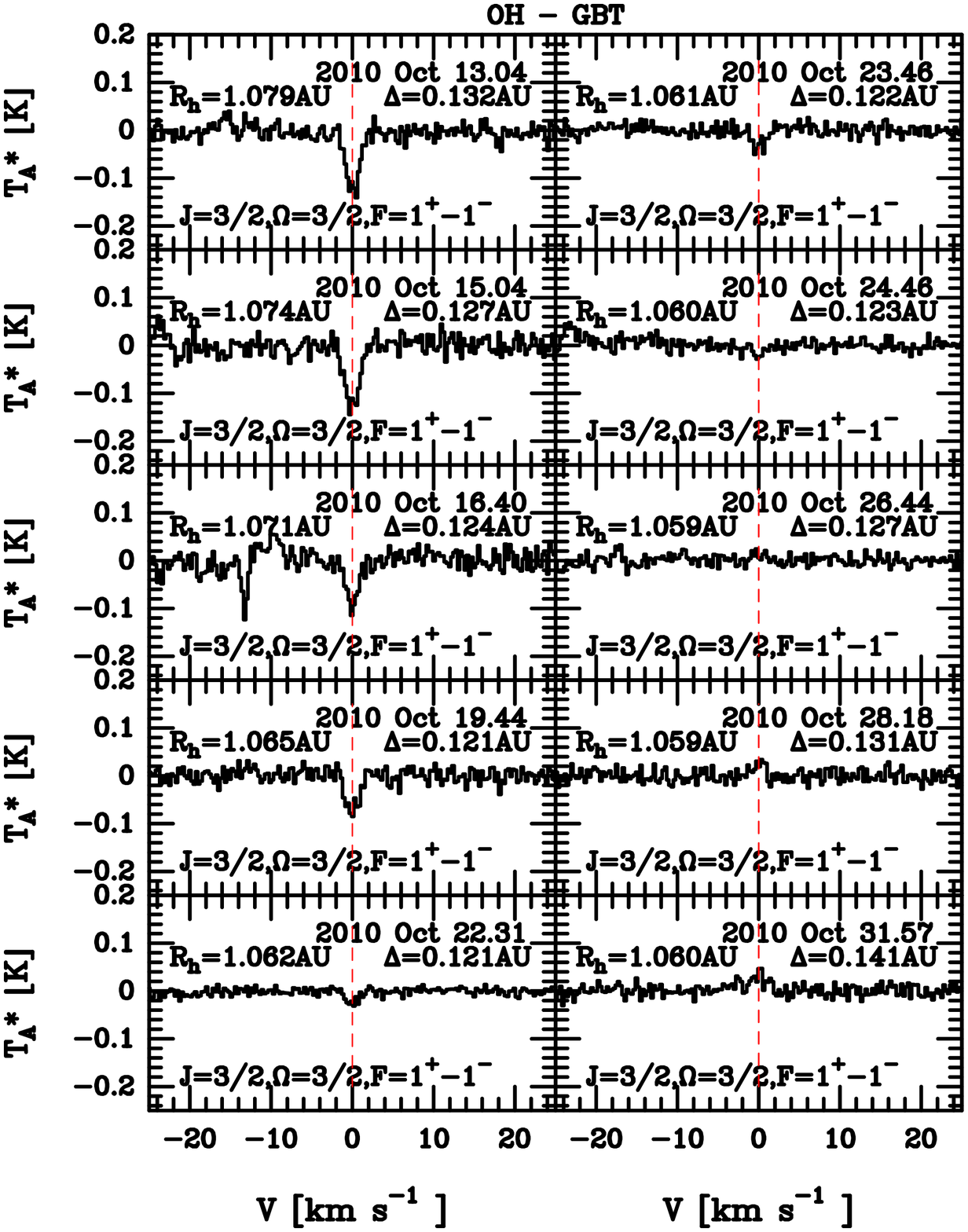}
\caption{Detection and upper limit of OH taken for each day of the observations with the GBT. The spectral resolution is $\approx$ 0.274 km $\rm s^{-1}$. Spectra are plotted in a cometocentric velocity frame. The red-dashed line denotes the J=3/2, $\Omega$=3/2, F=$\rm 1^+$-$\rm 1^-$ transition at the comet velocity.  Other features present in spectra are attributed to interstellar features at the time of observations.}
 \label{GBT_OH_APP}
\end{figure}

\section{Detection of HCN}\
\label{Appendix_HCN}
\begin{figure}[h!]
  \centering
\includegraphics[scale=0.7]{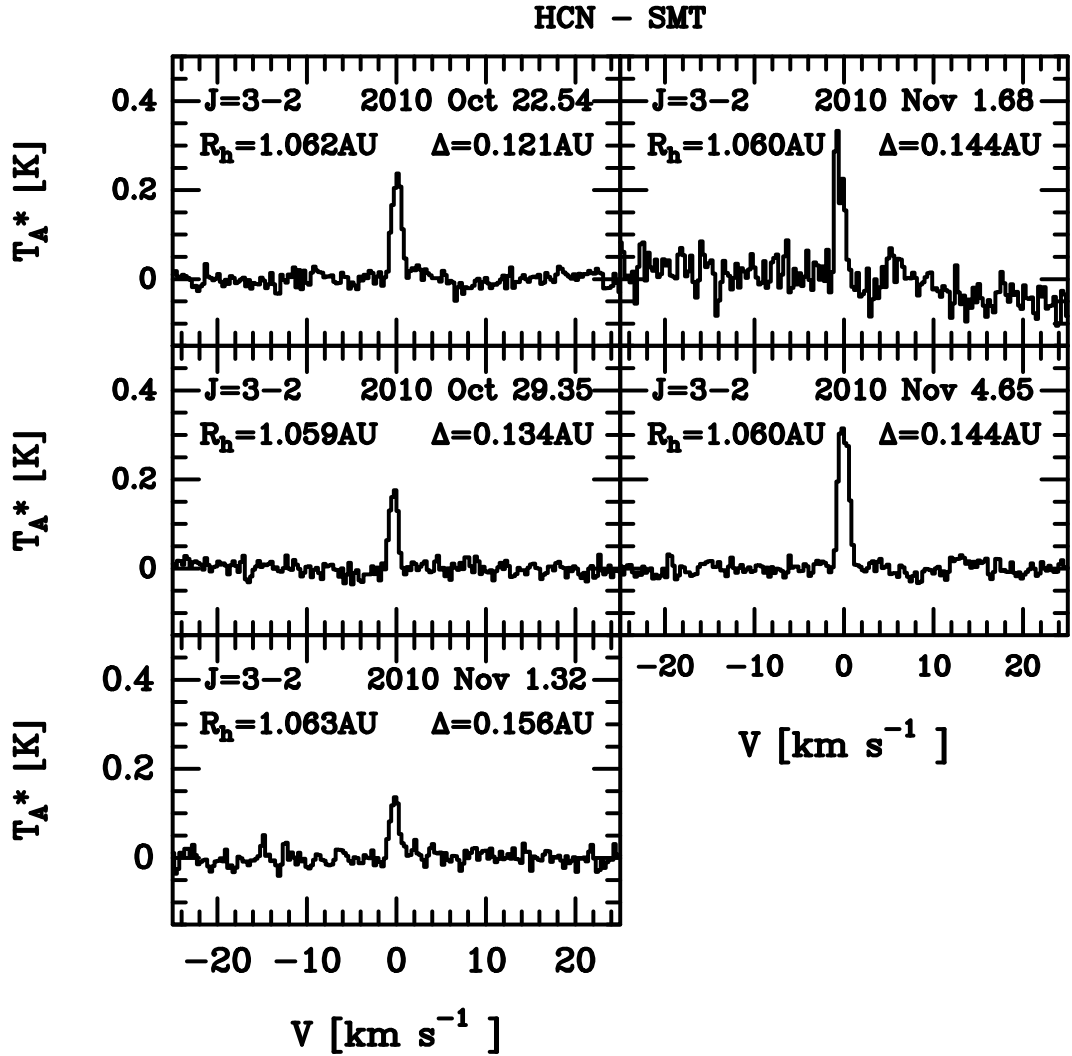}
\caption{Detection of HCN 3-2 taken for each day of observations with the SMT. The spectral resolution is $\approx$ 0.280 km $\rm s^{-1}$. Spectra are plotted in a cometocentric velocity frame.  }
 \label{HCN3-2_SMT}
\end{figure}

\begin{figure}[h!]
  \centering
\includegraphics[scale=0.5]{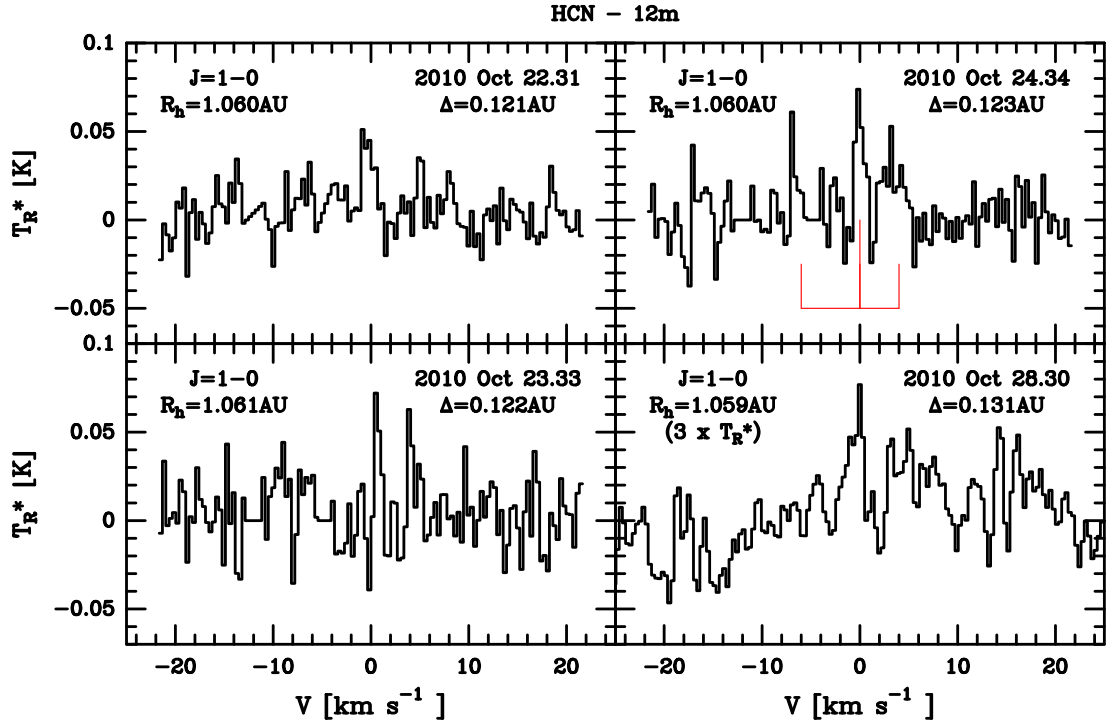}
\caption{Detection of HCN 1-0 for each day of observation with the 12m. The spectral resolution is $\approx$ 0.340 km $\rm s^{-1}$. Spectra are plotted in a cometocentric velocity frame. Relative intensities of the hyperfine lines are plotted below on 2010 Oct. 24.34. $T_R^{*}$ was multiplied by a factor 3 on Oct. 28.30.}
 \label{HCN1-0_12m}
\end{figure}

\begin{figure}[h!]
  \centering
\includegraphics[scale=0.3]{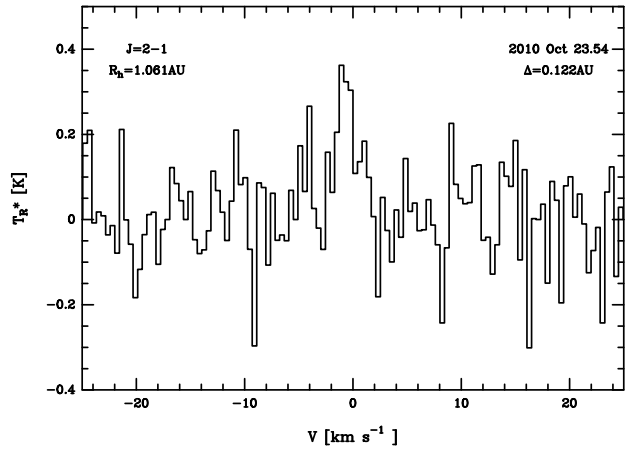}
\caption{Detection of HCN 2-1 taken on 23 October 2010 with the 12m. The spectral resolution is $\approx$ 0.170 km $\rm s^{-1}$. Spectra is plotted in a cometocentric velocity frame.}
 \label{HCN2-1_12m}
\end{figure}

\begin{figure}[h!]
  \centering
\includegraphics[scale=0.75]{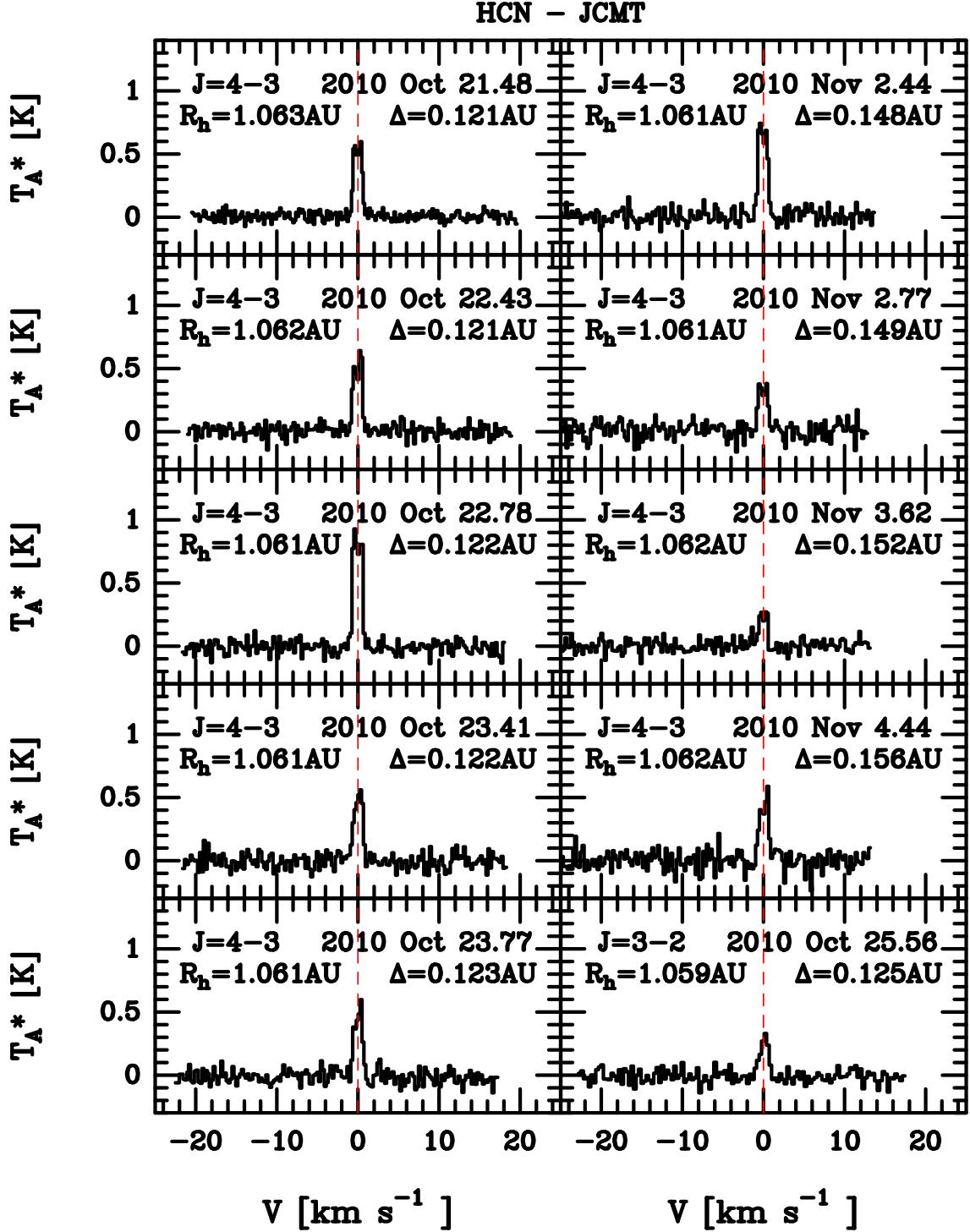}
\caption{Detection of HCN 4-3 and 3-2 taken for each day of observations with the JCMT. The spectral resolution are $\approx$ 0.232 km $\rm s^{-1}$ and $\approx$ 0.310 km $\rm s^{-1}$ for the 4-3 and 3-2 transition, respectively. Spectra is plotted in a cometocentric velocity frame. The observed asymmetry in the line profile observed can be attributed to localized outgassing as mentioned by \cite{Kawakita_2013}}.
 \label{HCN_JCMT}
\end{figure}

 \clearpage

\section{Detection of $\rm H_2CO$ }
\label{Appendix_H2CO}
\begin{figure}[h!]
  \centering
\includegraphics[scale=0.3]{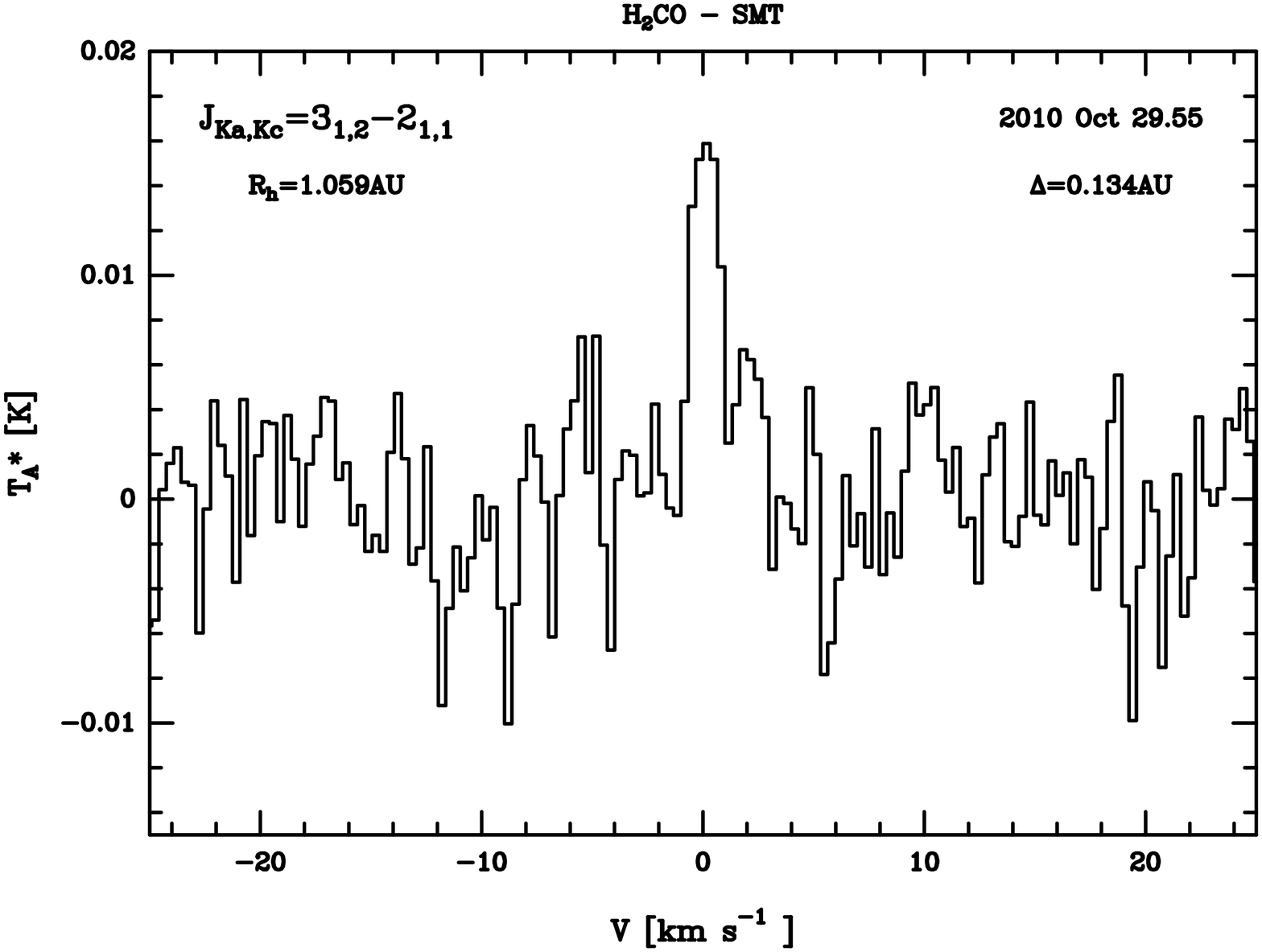}
\caption{Detection of $\rm H_2CO$ taken on 29 October 2010 with the SMT. The spectral resolution is $\approx$ 0.332 km $\rm s^{-1}$. Spectra are plotted in a cometocentric velocity frame.}
 \label{H2CO_SMT}
\end{figure}

\begin{figure}[h!]
  \centering
\includegraphics[scale=0.5]{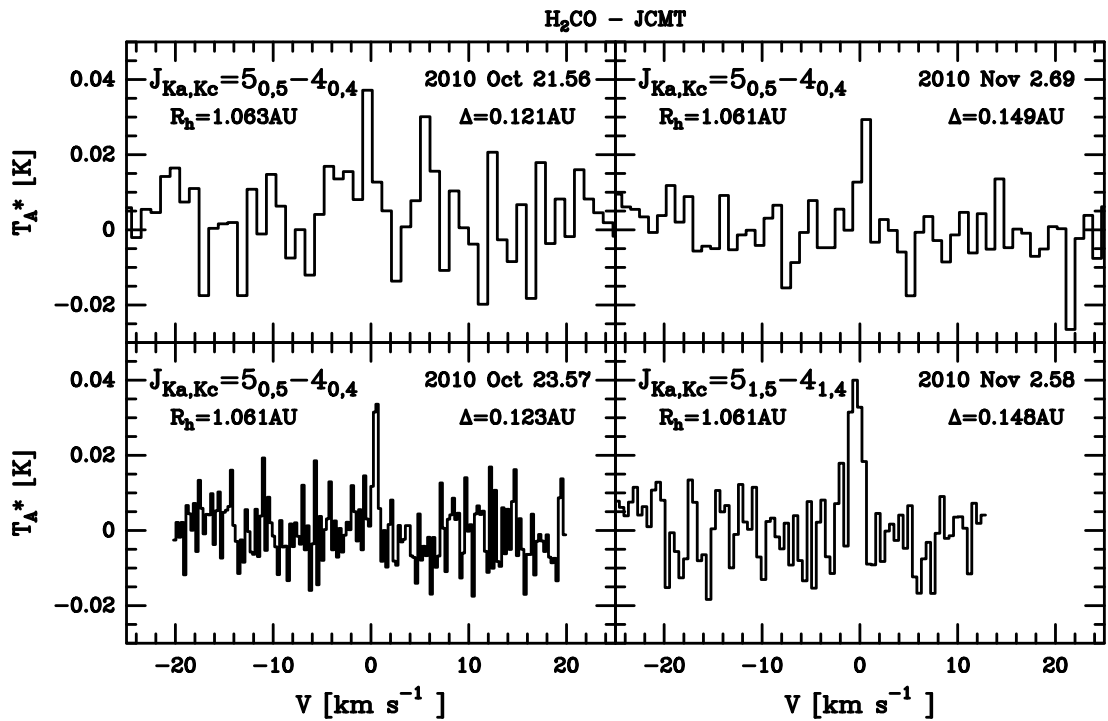}
\caption{Detection of $\rm H_2CO$ taken for each day of observations with the JCMT. The spectral resolution are $\approx$ 0.984 km $\rm s^{-1}$, $\approx$ 0.252 km $\rm s^{-1}$, $\approx$ 0.909 km $\rm s^{-1}$ and $\approx$ 0.468 km $\rm s^{-1}$ on 21 October, 23 October, 2.69 November and 2.58 November, respectively. Spectra are plotted in a cometocentric velocity frame.}
 \label{H2CO_JCMT}
\end{figure}
\clearpage

\section{Detection of $\rm HCN$ and upper limit of DCN }
\label{Appendix_DCN}
\begin{figure}[h!]
  \centering
\includegraphics[scale=0.4]{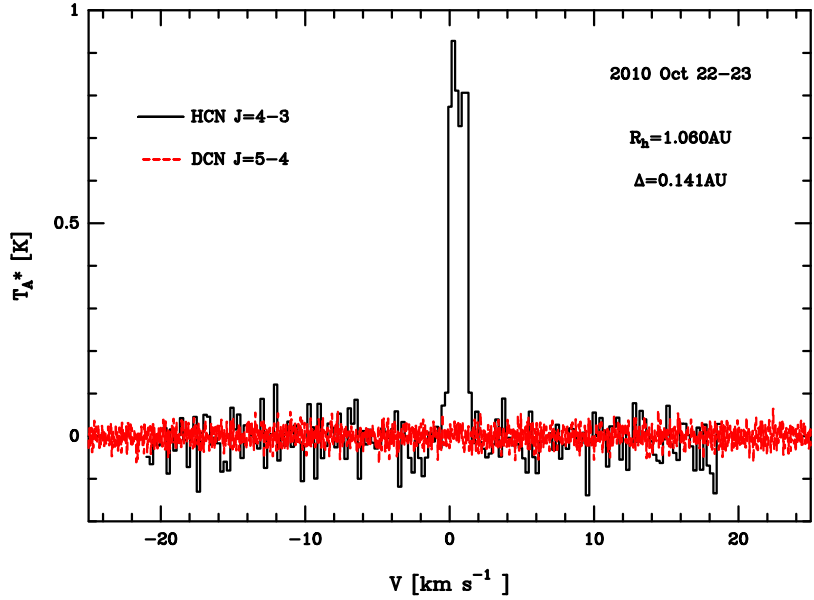}
\caption{Detection of HCN on taken on 22 October 2010 and average for the upper limit of DCN taken on 22-23 October 2010 with the JCMT. The spectral resolution are $\approx$ 0.232 km $\rm s^{-1}$ and $\approx$ 0.025 km $\rm s^{-1}$ for HCN and DCN, respectively. Spectra are plotted in a cometocentric velocity frame. The D/H ratio obtained here is $<$ 0.01}
 \label{DCN_data}
\end{figure}

\end{appendix}

\end{document}